\documentclass{aa}  

\usepackage{graphicx}
\usepackage{txfonts}
\usepackage{amsmath}    
\usepackage{amssymb}    
\usepackage{xspace}
\usepackage{booktabs}
\usepackage{threeparttable}
\usepackage{arydshln}
\usepackage{multicol}
\usepackage{subcaption}

\newcommand{\eref}[1]{Equation~\ref{#1}}
\newcommand{\fref}[1]{Figure~\ref{#1}}
\newcommand{\tref}[1]{Table~\ref{#1}}
\newcommand{\sref}[1]{Section~\ref{#1}}

\newcommand{\ie}{i.e.\@\xspace} 
\newcommand{\eg}{e.g.\@\xspace} 

\newcommand{\tess}{TESS\xspace}
\newcommand{\kepler}{{\it Kepler}\xspace}
\newcommand{\gaia}{{\it Gaia}\xspace}

\newcommand{\gamcep}{$\gamma$~Cep\xspace}
\newcommand{\planet}{$\gamma$~Cep~Ab\xspace}
\newcommand{\gamcepa}{$\gamma$~Cep~A\xspace}
\newcommand{\gamcepb}{$\gamma$~Cep~B\xspace}

\newcommand{\vsini}{$v \sin i_{\rm A}$\xspace}
\newcommand{\teff}{$T_{\rm eff}$\xspace}
\newcommand{\feh}{$\rm [Fe/H]$\xspace}
\newcommand{\logg}{$\log g$\xspace}
\newcommand{\numax}{\ensuremath{\nu_{\rm max}}\xspace}
\newcommand{\dnu}{\ensuremath{\Delta\nu}\xspace}

\newcommand{\finalmass}{$1.27^{+0.05}_{-0.07}$~M$_\odot$\xspace}
\newcommand{\finalradius}{$4.74^{+0.07}_{-0.08}$~R$_\odot$\xspace}
\newcommand{\finalage}{$5.7^{+0.8}_{-0.9}$~Gyr\xspace}

\newcommand{\finalnumax}{$191^{+3}_{-4}$~$\mu$Hz\xspace}
\newcommand{\finaldnu}{$14.59^{+0.06}_{-0.05}$~$\mu$Hz\xspace}

\newcommand{\finalmtwo}{$0.328^{+0.009}_{-0.012}$~M$_\odot$\xspace}
\newcommand{\finalmp}{$6.6^{+2.3}_{-2.8}$~M$_{\rm Jup}$\xspace}

\newcommand{\RomanNumeralCaps}[1]
    {\MakeUppercase{\romannumeral #1}}

\begin{document}

   \title{Solar-like oscillations in $\gamma$ Cephei A as seen through SONG and TESS}
   \subtitle{A seismic study of $\gamma$ Cephei A}

   \author{
   E.~Knudstrup\inst{\ref{sac},\ref{not}} \and
    M.~N.~Lund\inst{\ref{sac}} \and
    M.~Fredslund~Andersen\inst{\ref{sac}} \and
    J.~L.~R\o rsted\inst{\ref{sac}} \and
    F.~P\'{e}rez~Hern\'{a}ndez\inst{\ref{iac},\ref{ull}} \and
    F.~Grundahl\inst{\ref{sac}} \and
    P.~L.~Pall\'{e}\inst{\ref{iac},\ref{ull}} \and
    D.~Stello\inst{\ref{syd},\ref{sifa},\ref{sac}} \and
    T.~R.~White\inst{\ref{sifa}} \and
    H.~Kjeldsen\inst{\ref{sac}} \and
    M.~Vrard\inst{\ref{ohio}} \and
    M.~L.~Winther\inst{\ref{sac}} \and
    R.~Handberg\inst{\ref{sac}} \and
    S.~Sim\'{o}n-D\'{i}az\inst{\ref{iac},\ref{ull}}
  }

   \authorrunning{Knudstrup et~al.}
   \institute{
    Stellar Astrophysics Centre, Department of Physics and Astronomy, Aarhus University, Ny Munkegade 120, DK-8000 Aarhus C, Denmark\label{sac} \and
    Nordic Optical Telescope, Rambla José Ana Fernández Pérez 7, E-38711 Breña Baja, Spain\label{not} \and
    Instituto de Astrofísica de Canarias. E-38205 La Laguna, Tenerife, Spain\label{iac} \and
    Universidad de La Laguna (ULL), Departamento de Astrofísica, E-38206 La Laguna, Tenerife, Spain\label{ull} \and
    School of Physics, The University of New South Wales, Sydney NSW 2052, Australia\label{syd} \and
    Sydney Institute for Astronomy (SIfA), School of Physics, University of Sydney, Camperdown, NSW 2006, Australia\label{sifa} \and
    Department of Astronomy, The Ohio State University, Columbus, OH 43210, USA\label{ohio}
   }

   \date{Received ...; accepted ...}

 
  \abstract
  {Fundamental stellar parameters such as mass and radius are some of the most important building blocks in astronomy, both when it comes to understanding the star itself and when deriving the properties of any exoplanet(s) they may host. Asteroseismology of solar-like oscillations allows us to determine these parameters with high precision.}
  {We investigate the solar-like oscillations of the red-giant-branch star \gamcepa, which harbours a giant planet on a wide orbit.} 
  {We did this by utilising both ground-based radial velocities from the SONG network and space-borne photometry from the NASA \tess mission.} 
  {From the radial velocities and photometric observations, we created a combined power spectrum, which we used in an asteroseismic analysis to extract individual frequencies. We clearly identify several radial and quadrupole modes as well as multiple mixed, dipole modes. We used these frequencies along with spectroscopic and astrometric constraints to model the star, and we find a mass of \finalmass, a radius of \finalradius, and an age of \finalage. We then used the mass of \gamcepa and our SONG radial velocities to derive masses for \gamcepb and \planet of \finalmtwo and \finalmp, respectively.} 
  {}


   \maketitle
%


\section{Introduction}

Owing to its brightness, the Gamma Cephei (\gamcep) system has a long and rich history in the astronomical literature. In their high-precision radial velocity (RV) survey for planetary companions around nearby, stars \citet{Campbell1988} found \gamcep to be a single-lined spectroscopic binary. On top of the large-scale RV signal induced by the binary orbit of the secondary companion, \citet{Campbell1988} found an additional variation with a periodicity of around 2.7~yr and an amplitude of some 25~m~s$^{-1}$, which they suspected to be due to a third body in the system orbiting the primary star. As such, \gamcep was amongst one of the first systems proposed to harbour an extra-solar planet (exoplanet). The planetary nature of this third body, \planet or Tadmor\footnote{Following the 2015 edition of the IAU NameExoWorlds initiative (\url{https://www.iau.org/news/pressreleases/detail/iau1514/}).}, was confirmed by \citet[][]{Hatzes2003}, who found a period of around 906~d and an amplitude of 27.5~m~s$^{-1}$, corresponding to a minimum mass of 1.7~M$_{\rm J}$.

In \citet[][]{Reffert2011} the astrometric orbit of \gamcep was investigated using {\it Hipparcos} data \citep[][]{vanleeuwen2007}. This enabled the authors to place constraints on the orbital inclination of the planet, finding minimum and maximum values of $i=3.7^\circ$ and $i=15.5^\circ$, corresponding, respectively, to $28.1$~M$_{\rm J}$ and $6.6$~M$_{\rm J}$. The mass range for \planet of  ${\sim}13-80$~M$_{\rm J}$ thus straddles the border between planetary and brown dwarf regimes \citep[][]{Baraffe2002,Spiegel2011}. \citet[][]{Neuhauser2007} found an orbital inclination for the binary orbit of $i_{\rm AB}=119.3 \pm 1.0 ^\circ$, meaning that the planetary orbit is perpendicular to the binary orbit. In addition, they determined the masses of the stars to be $M_{\rm A} = 1.40 \pm 0.12$~M$_\odot$ for the primary and $M_{\rm B} = 0.409 \pm 0.018$~M$_\odot$ for the secondary. The primary component is thus consistent with being a `retired' A star, while the secondary is an M dwarf.

A stellar fly-by after the formation of the planet has been suggested as being responsible for tilting the binary orbit \citep[][]{Marti2012}. Recently, \citet[][]{Huang2022} employed the eccentric Kozai-Lidov mechanism \citep[][]{Kozai1962,Lidov1962}  to explain the high mutual inclination. The exact orbital configurations and the masses involved have significant consequences for our understanding of how the system might have formed and since evolved. For instance, \citet[][]{Jang-Condell2008} argue that \gamcepb should have truncated the protoplanetary disk around \gamcepa, which could have limited planet formation in the disk.

Clearly, \gamcep is an intriguing system in the contexts of planet formation, dynamical evolution, and system architectures. Understanding these processes requires intricate knowledge of the fundamental properties of the host star. Asteroseismology is an important tool in stellar characterisation, directly linking the observed oscillation frequencies to stellar properties such as mass, radius, and age \citep[][]{Aerts2010}. Solar-like oscillations are highly prevalent in subgiant and red-giant-branch (RGB) stars with amplitudes of several hundred ppm as observed in photometry \citep[\eg][]{Huber2019,Stokholm2019,Li2020} and several m~s$^{-1}$ in RV \citep[\eg][]{Stello2017}, which can easily be detected with modern photometers and spectrographs. With periods of a few hours, it is not only possible, but also straightforward, to carry out asteroseismic studies of RGB stars with ground-based facilities \citep[see \eg][]{Grundahl2017, Frandsen2018}.

While the mass of \gamcepa would suggest it was originally an A-type star, it has long been known to be an evolved star \citep[e.g.][]{Eggen1955} and has since transitioned into a K-type star, making it a viable target for asteroseismic studies of solar-like oscillations. The oscillations in \gamcepa were investigated in \citet[][]{Stello2017} in a study of the retired A-star controversy, which refers to the fact that there appears to be an overabundance of relatively massive planet-hosting stars \citep[][]{Johnson2014,North2017,Campante2017,Hjorringgaard2017}. For this investigation they made use of observations from the Stellar Observations Network Group (SONG) project \citep[][]{Grundahl2017}. This was further investigated with additional SONG observations in \citet[][]{Malla2020}, who found a mass of $1.32 \pm 0.12$~M$_\odot$ from the average seismic parameters: the large frequency separation, \dnu, and the frequency of maximum power, \numax.

Here we expand upon the asteroseismic and orbital analysis of \gamcepa through additional ground-based observations from the SONG network as well as space-based photometry from the Transiting Exoplanet Survey Satellite \citep[TESS;][]{Ricker2015} mission. The paper is structured as follows: in \sref{sec:obs} we present the observations and spectroscopic analysis. Our asteroseismic data analysis is detailed in \sref{sec:anal}, and our modelling is described in \sref{sec:model}. We discuss our results in \sref{sec:disc} and conclude in \sref{sec:conc}.

\section{Observations}\label{sec:obs}
To detect oscillations of the primary star in the system, \gamcep has been closely monitored with both the 1 m fully robotic Hertzsprung SONG telescope \citep{Fredslundandersen2019,Grundahl2017} on Observatorio del Teide, Tenerife, Spain, and the Chinese SONG node \citep{Deng2013} at the Delingha Observing Station, China. In addition to the ground-based data \gamcep has also been observed by TESS in three sectors. All the timestamps from both SONG nodes and TESS are shown in \fref{fig:campaigns}, clearly showing the overlap between the 2019 SONG campaign and the Sector 18 TESS observations. In addition to the SONG and TESS data, we observed \gamcep with the Nordic Optical Telescope at the Roque de los Muchachos, La Palma, Spain, using the       FIber-fed Echelle Spectrograph \citep[FIES;][]{Frandsen1999,Telting2014}.

\subsection{SONG data}
\label{sec:song_data} 
\begin{figure*}
    \centering
        \includegraphics[width=\textwidth]{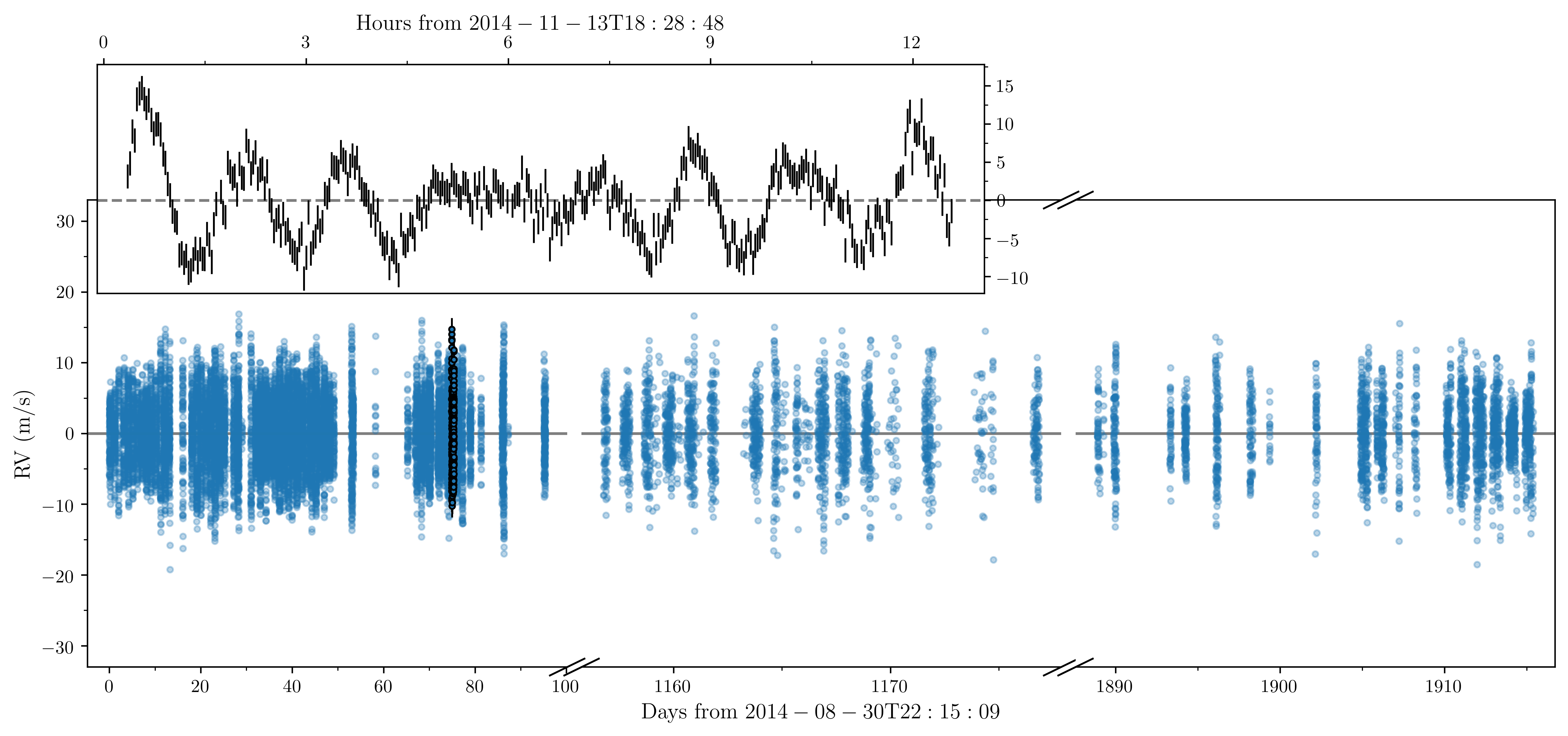}
    \caption{Three different seismic SONG campaigns of \gamcep after filtering and subtracting a nightly median (\sref{sec:filter}). The error bars in the inset represent one night of observations from the 2014 campaign. The data from the inset are displayed in the main plot as blue circles with black edges and error bars.}
    \label{fig:song_rv}
\end{figure*}

The SONG data were obtained under the programmes P00-02 (P.I. Pere Pallé, IAC), P06-06 (P.I. Dennis Stello, UNSW), and P10-01 (P.I. Mads Fredslund Andersen, AU) during Autumn of 2014, 2017, and 2019. The observations were carried out with an iodine cell for precise wavelength calibration, an exposure time of 120 seconds, and using slit number 6, which corresponds to a resolving power $(\lambda/\Delta\lambda)$ of 90,000. Each 1D spectrum covering the wavelength range from 4400 to 6900 \AA~was extracted and the RVs obtained following the procedures outlined in \citet{Grundahl2017} using the \texttt{iSONG} reduction code \citep{Corsaro2012a,Antoci2013}. All RVs were obtained using the same high resolution template to ensure no unwanted shifts in RVs were introduced between the different datasets. In \fref{fig:song_rv} we show the RVs from the three SONG campaigns. The inset shows one night of observations in the 2014 campaign with  error bars. In \tref{tab:campaigns} we summarise the different observing campaigns. 

\begin{table}[]
    \centering
    \caption{{\bf SONG campaigns.}}
    \begin{threeparttable}
    \begin{tabular}{c c c c c}
        \toprule
         Node & Start & End & $N$ & $\Delta t$  \\
          & (dd-mm-yyyy) & (dd-mm-yyyy) & & (s) \\ 
        \midrule
         Tenerife & 30-08-2014 & 14-11-2014 & 12666 & 120 \\
         Tenerife & 30-10-2017 & 10-11-2017 & 837 & 120 \\
         Delingha & 30-10-2017 & 10-11-2017 & 2083 & 180 \\
         Tenerife & 01-11-2019 & 28-11-2019 & 2573 & 120 \\
         \bottomrule
    \end{tabular}
    \begin{tablenotes}
        \item The start and end date of a SONG campaign at a given node with number of observations ($N$) and the exposure time ($\Delta t$).
    \end{tablenotes}
    \end{threeparttable}
    \label{tab:campaigns}
\end{table}

\subsection{TESS data}
\label{sec:tess_data}
The \gamcep system was observed by TESS in Sectors 18 (November 2019), 24-25 (mid-April to the beginning of June 2020), and 52 (mid-May to mid-June 2022). In all four sectors \gamcep was observed in TESS' 2-minute cadence mode as summarised in \tref{tab:sectors}. We downloaded the extracted TESS light curves of \gamcep from the Mikulski Archive for Space Telescopes (MAST) created by the Science Processing Operations Center \citep[SPOC;][]{Jenkins2016}, which uses Simple Aperture Photometry \citep[SAP;][]{Twicken2010,book:morris2017}. Common instrumental systematics were removed through the Presearch Data Conditioning \citep[PDCSAP;][]{Smith2012,Stumpe2012} algorithm, and we used these PDCSAP light curves in our analysis. We also tested a photometric extraction using the $\rm K2P^2$ pipeline \citep[][]{Lund2015} with custom apertures but with a similar quality to the PDCSAP data. We chose to adopt the PDCSAP method for the sake of better reproducibility. 

\begin{figure*}
    \centering
        \includegraphics[width=\textwidth]{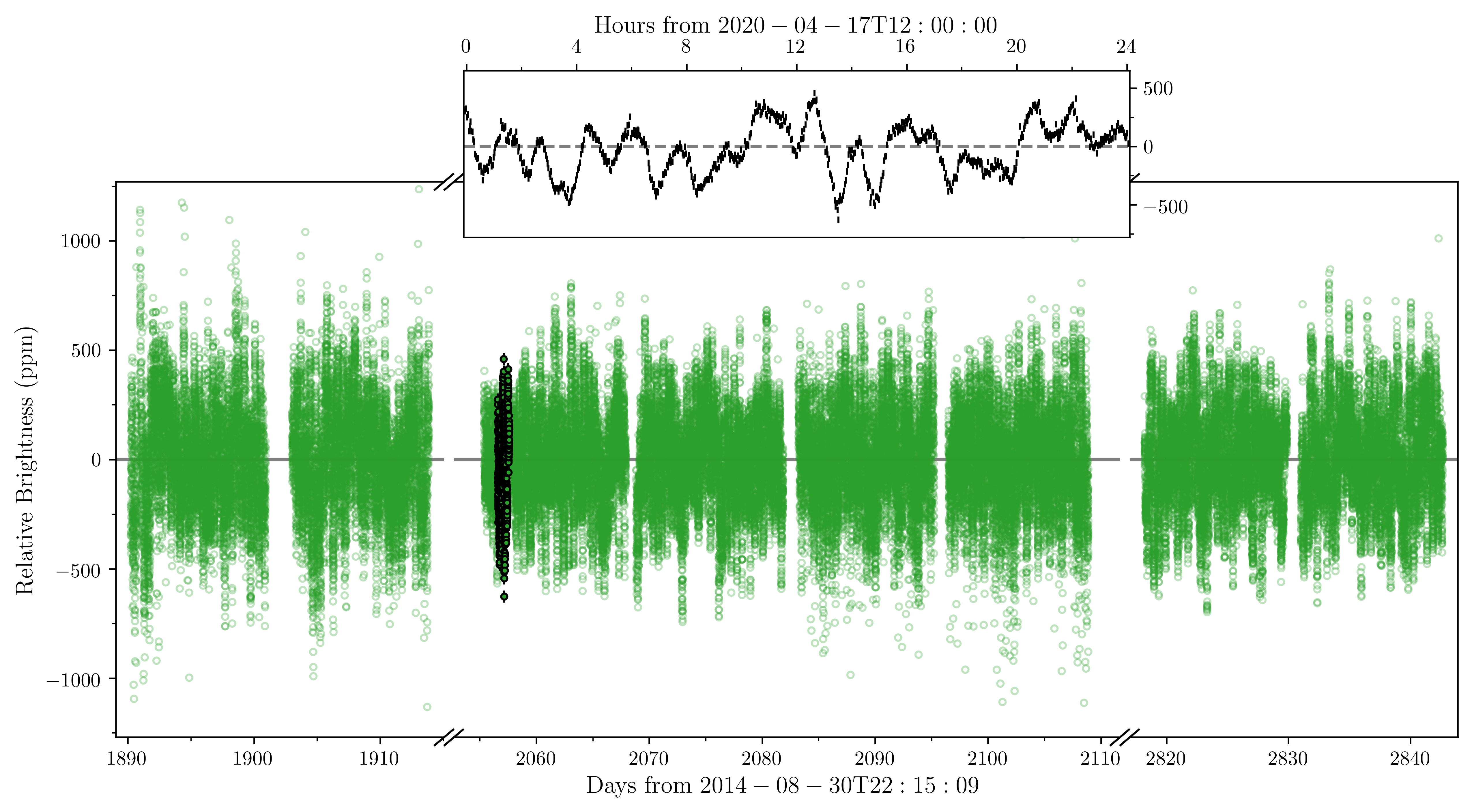}
    \caption{Light curve of \gamcep as observed by TESS in Sectors 18, 24, 25, and 52. Here we have removed outliers and normalised the light curve as described in \sref{sec:filter}. The error bars in the inset show a 24~hr interval. The data shown in the inset are highlighted in the main plot as green circles with black edges and error bars.}
    \label{fig:tess_lc}
\end{figure*}

\begin{table}[]
    \centering
    \caption{{\bf TESS sectors.}}
    \begin{threeparttable}
    \begin{tabular}{c c c c c}
        \toprule
         Sector & Start & End & $N$ & $\Delta t$  \\
          & (dd-mm-yyyy) & (dd-mm-yyyy) & & (s) \\ 
        \midrule
         18 & 02-11-2019 & 27-11-2019 & 15135 & 120 \\
         24 & 16-04-2020 & 13-05-2020 & 18216 & 120 \\
         25 & 13-05-2020 & 08-06-2020 & 17238 & 180 \\
         52 & 18-05-2022 & 13-06-2022 & 16748 & 120 \\
         \bottomrule
    \end{tabular}
    \begin{tablenotes}
        \item The start and end date of a TESS Sector along with the number of exposures ($N$, unflagged cadence) and the cadence ($\Delta t$).
    \end{tablenotes}
    \end{threeparttable}
    \label{tab:sectors}
\end{table}

\subsection{Spectroscopic analysis}

The observations of \gamcep using FIES (carried out in 2021) were obtained to get high signal-to-noise (S/N) spectra to extract spectroscopic parameters. For the spectral analysis we used all our FIES spectra with the programme \texttt{iSpec} \citep{Blanco-Cuaresma2014,Blanco-Cuaresma2019}. First, we normalised the spectra by fitting the continuum using splines. We then calculated the RV of the star at each epoch and shifted the spectra into the rest frame. To derive the stellar parameters we used \texttt{iSpec} with the code \texttt{SPECTRUM} \citep{Gray1994}, which creates a synthetic stellar spectrum to compare against our observed spectra. We opted for the MARCS \citep{Gustafsson2008} grid of model atmospheres as the template for the synthetic spectrum. 

We followed the approach described in \citet{Lund2016} in which the spectroscopic parameters are determined through an iterative process due to degeneracies in the estimates of \teff (the effective temperature), \logg (surface gravity), and \feh \citep[metallicity;][]{Smalley2005,Torres2012}. We used our measured value of \numax of $185.6$~$\mu$Hz (\tref{tab:seis_model}) with the \teff from \citet{Mortier2013} of $4764 \pm 122$~K as an initial value to estimate the seismic \logg as

\begin{equation}
    g \simeq g_\odot \left ( \frac{\nu_{\rm max}}{\nu_{\rm max,\odot}} \right ) \left ( \frac{T_{\rm eff}}{T_{\rm eff,\odot}} \right )^{1/2} \, ,
    \label{eq:g}
\end{equation}
with $\nu_{\rm max,\odot}=3090\pm30$~$\mu$Hz, $T_{\rm eff,\odot}=5777$~K, and $g_\odot=27,402$~cm~s$^{-1}$ \citep{Brown1991,Kjeldsen1995,Huber2011,Chaplin2014}.
Initially, we used those values as starting values in a fit for each epoch where all parameters were free to vary. 
From this initial fit we got $T_{\rm eff} = 5094$~K as the median for all epochs, which yields a \logg of 3.19 (from \eref{eq:g}). In the second iteration we thus fixed \logg at 3.19, while letting $T_{\rm eff}$, $\rm [Fe/H]$, \vsini (projected rotation speed), $\zeta$ (macro-turbulence), and $\xi$ (micro-turbulence) free to vary. The resulting median \teff across epochs was $4806$~K, which gives a \logg of 3.18. We thus considered the fit to have converged.

\begin{figure}
    \centering
        \includegraphics[width=\columnwidth]{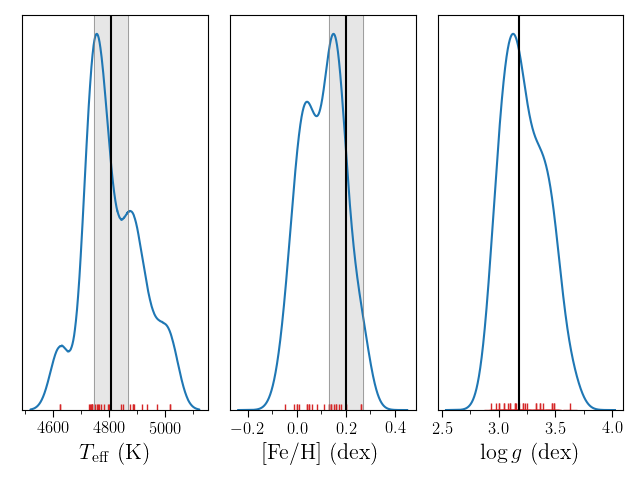}
    \caption{{\bf Literature comparison.} The panels show the KDEs of the values for \teff, \feh, and \logg obtained from the literature after year 2000 via the \texttt{SIMBAD} database. The reported values of the individual studies are indicated with red markers. The vertical solid lines and shaded regions show our values and associated uncertainties. }
    \label{fig:lit_comp}
\end{figure}

We find the results to be very consistent from epoch to epoch, which leads to very precise measurements of the parameters. 
These do not account for any systematic uncertainties, which are undoubtedly present. We therefore follow the approach of \citet{Torres2012} and add (in quadrature) uncertainties for \teff, \feh, and \vsini of 59~K, 0.062~dex, and 0.85~km~s$^{-1}$, respectively. The results are summarised in \tref{tab:star}.
We also find that our spectroscopic results agree well with the range of values found in the literature -- \fref{fig:lit_comp} shows the kernel density estimations (KDEs) of results for \teff, \feh, and \logg from the literature after year 2000, as collected from the \texttt{SIMBAD} database \citep[][]{Wenger2000}.

\begin{table}
    \centering
    \caption{ {\bf System parameters.} }
    \begin{threeparttable}
    \begin{tabular}{c c c}
        \toprule 
        \multicolumn{3}{c}{\gamcepa} \\
        \multicolumn{3}{c}{} \\
        Parameter & Value & Description \\
        \midrule
         TIC\tnote{a} & 367912480  & \\
         HD & 222404 & \\
         \midrule
         $\alpha$ (J2000)\tnote{b} & 23 39 20.59 & Right ascension (R.A.) \\
         $\delta$ (J2000)\tnote{b} & +77 37 59.25 & Declination (Dec.) \\
         $\mu_\alpha$ (mas~yr$^{-1}$)\tnote{b} & $-64.86 \pm 0.14$ & Proper motion R.A. \\
         $\mu_\delta$ (mas~yr$^{-1}$)\tnote{b} & $171.16 \pm 0.14$ & Proper motion Dec. \\
         $\varpi$ (mas)\tnote{b} & $72.52 \pm 0.15$ & Parallax \\
         dist (pc)\tnote{d} & $13.78\pm 0.03$ & Distance \\
         \midrule
         $G$\tnote{b} & $2.9456\pm0.031^*$ & {\it Gaia} $G$ magnitude\\
         $L_G$ (L$_\odot$) & $10.48\pm 0.23$ & Lum. {\it Gaia} $G$-band\\
         $V$\tnote{c} & $3.212\pm 0.008$ & Tycho $V$ magnitude\\
         $B-V$\tnote{c} & $1.028\pm 0.004$ & Tycho colour\\
         \midrule
         SpT\tnote{e} & K1\RomanNumeralCaps{3}-\RomanNumeralCaps{4} & Spectral type \\
         \midrule
          $T_{\rm eff}$ (K)\tnote{f} & $4806 \pm 60$ & Effective temperature \\
           $\log g$ (cgs; dex)\tnote{f} & $3.18$ & Surface gravity \\
          $\rm [Fe/H]$ (dex)\tnote{f} & $0.20 \pm 0.07$ & Metallicity \\      

          \vsini (km~s$^{-1}$)\tnote{f} & $0.0 \pm 0.9$ & Projected rotation \\
          $\zeta$ (km~s$^{-1}$)\tnote{f} & $3.77 \pm 0.04$ & Macro-turbulence \\
          $\xi$ (km~s$^{-1}$)\tnote{f} & $1.14 \pm 0.02$ & Micro-turbulence \\

         \bottomrule
    \end{tabular}
\begin{tablenotes}
    \item Catalogue IDs, coordinates, magnitudes, spectral type, and spectroscopic parameters.
    \item[a] \url{https://exofop.ipac.caltech.edu/tess/}.
    \item[b] \citet{GaiaDR3}.
    \item[c] \citet{Mermilliod1997}.
    \item[d] \citet{BailerJones2021} (photogeometric).
    \item[e] \citet{Keenan1989}.
    \item[f] This work from FIES spectra using iSpec.
    \item[*] Corrected following \citet{Riello2021}.
\end{tablenotes}
\end{threeparttable}
    \label{tab:star}
\end{table}

\subsection{Luminosity from \gaia}
As an additional constraint for the seismic modelling we compute the stellar luminosity from combining a distance measure with the \gaia Data Release 3 (DR3) $G$-band magnitude following \citep[see \eg][]{Torres2010}:
\begin{equation}\label{eq:lum}
    L/L_{\odot} = 10^{0.4\left(5\log_{10}(d) - G + A_G - BC_G + V_{\odot} + 26.572 + BC_{V, \odot}\right)}\, ,
\end{equation}
where $d$ is the distance in pc, $G$ is the apparent \gaia Early Data Release 3 $G$-band magnitude, $A_G$ is the extinction in the $G$ band, and $BC_G$ is the bolometric correction. We use the photogeometric distance from \citet[][see \tref{tab:star}]{BailerJones2021}, adopt values of $V_{\odot}=-26.74\pm 0.01$ mag and $BC_{V, \odot}=-0.078\pm0.005$ mag from analysis of empirical solar spectra (Lund et al., in prep.), and make saturation corrections to the \gaia photometry following \citet{Riello2021}. For the bolometric correction $BC_G$ we use the interpolation routines of \citet{Casagrande2018}.
Based on the \emph{Stilism}\footnote{https://stilism.obspm.fr/} 3D reddening map \citep{Lallement2014,Capitanio2017,Lallement2019} we adopt a zero extinction, which is consistent with the close proximity of the system at a distance of only ${\sim}13.2$ pc.
The resulting luminosity is provided in \tref{tab:star}.

\section{Seismic analysis}\label{sec:anal}

\begin{table*}
    \centering
    \caption{{\bf Stellar parameters.}  
    }
    \begin{threeparttable}
    \begin{tabular}{c c c c c c c c c}
    \toprule
Source & Mass & Radius & $\tau$ & $T_{\rm eff}$ & $\log g$ & $\rm [Fe/H]$ & \numax & \dnu  \\ 
 & (M$_\odot$) & (R$_\odot$) & (Gyr) & (K) & (cgs; dex) & (dex) & ($\mu$HZ) & ($\mu$HZ) \\ 
\midrule 
\citet{Stello2017} & $1.32 \pm 0.20$ & $4.88 \pm 0.22$\tnote{a} & - & $4764 \pm 122$\tnote{b} & $3.17 \pm 0.08$ & $0.13 \pm 0.06$\tnote{b} & $185 \pm 28$ & -  \\ 
\citet{Malla2020} & $1.32 \pm 0.12$ & $4.88 \pm 0.22$ & - & - & - & - & $185 \pm 9$ & $14.28 \pm 0.58$  \\ 

SONG\tnote{c} &  -  &  -  &  -   & - &  -  &  - & $190.1 \pm 0.7$ & - \\ 
TESS\tnote{c} &  -  &  -  &  -   &  - &  -  &  - & $185.6^{+1.0}_{-0.9}$ & - \\ 
 
BASTA &  $ 1.27^{+0.05}_{-0.07} $  &  $ 4.74^{+0.07}_{-0.08} $  &  $ 5.7^{+0.8}_{-0.9} $  &  $ 4775^{+33}_{-31} $ &  $ 3.189^{+0.007}_{-0.008}$  &  $ 0.18\pm0.06 $ & $191^{+3}_{-4}$ & $14.59^{+0.06}_{-0.05}$ \\

\bottomrule
    \end{tabular}
\begin{tablenotes}
    \item Physical properties for \gamcepa derived from asteroseismology from the literature and in this work.
   
    \item[a] Derived from spectroscopic parameters in \citet{Mortier2013}.
    \item[b] From \citet{Mortier2013}.
    \item[c] From the full time series.

\end{tablenotes}
\end{threeparttable}

    \label{tab:seis_model}
\end{table*}

\subsection{Filtering}
\label{sec:filter}

For the asteroseismic analysis we filtered the SONG data on a night-by-night basis using a locally weighted scatterplot smoothing \citep[LOWESS;][]{misc:statsmodels2010} filter, which takes both the duration of the observations and the fill factor on a given night into account. Firstly, we filtered out the worst outliers by crudely removing all points deviating by more than $6 k$ times the median absolute deviation (MAD) and $k\approx1.4826$ (for normally distributed data). We then used the LOWESS filter to smooth the data, thereby removing the oscillations. We define the fill factor as $f \equiv N\delta t/\tau$ with $\delta t$ being the sampling (2 min.), $\tau$ the duration of the night, and $N$ the number of data points. For nights with $N<20$ or $f < 0.3$ we simply use the median to flatten the time series. In this flat(ter) time series we removed all points ${>3k \times \mathrm{MAD}}$. We furthermore use the nightly MAD as the uncertainties for the data acquired on that night. The resulting time series can be seen in \fref{fig:song_rv}. 

Equivalently, for the \tess data we used a LOWESS, but on a \tess orbit-to-orbit basis, that is, in intervals of around 13.7~d. Again, we started out by removing all data points with deviations of more than $6 k \times \mathrm{MAD}$, then applied the filter to flatten the time series, and removed the outliers, though using a more conservative rejection criterion and only removing points with a MAD of more than 6 as opposed to 3 for the RVs (as the SONG data were more prone to outliers, e.g. because of poor weather). To ensure that the uncertainties on the two datasets (RVs and photometry) were derived in a consistent way, we estimated the uncertainties as the daily (24~hr) MAD as shown in \fref{fig:tess_lc}.

\subsection{Power spectra}
\label{sec:spectra}
\begin{figure*}
    \centering
    \includegraphics[width=\textwidth]{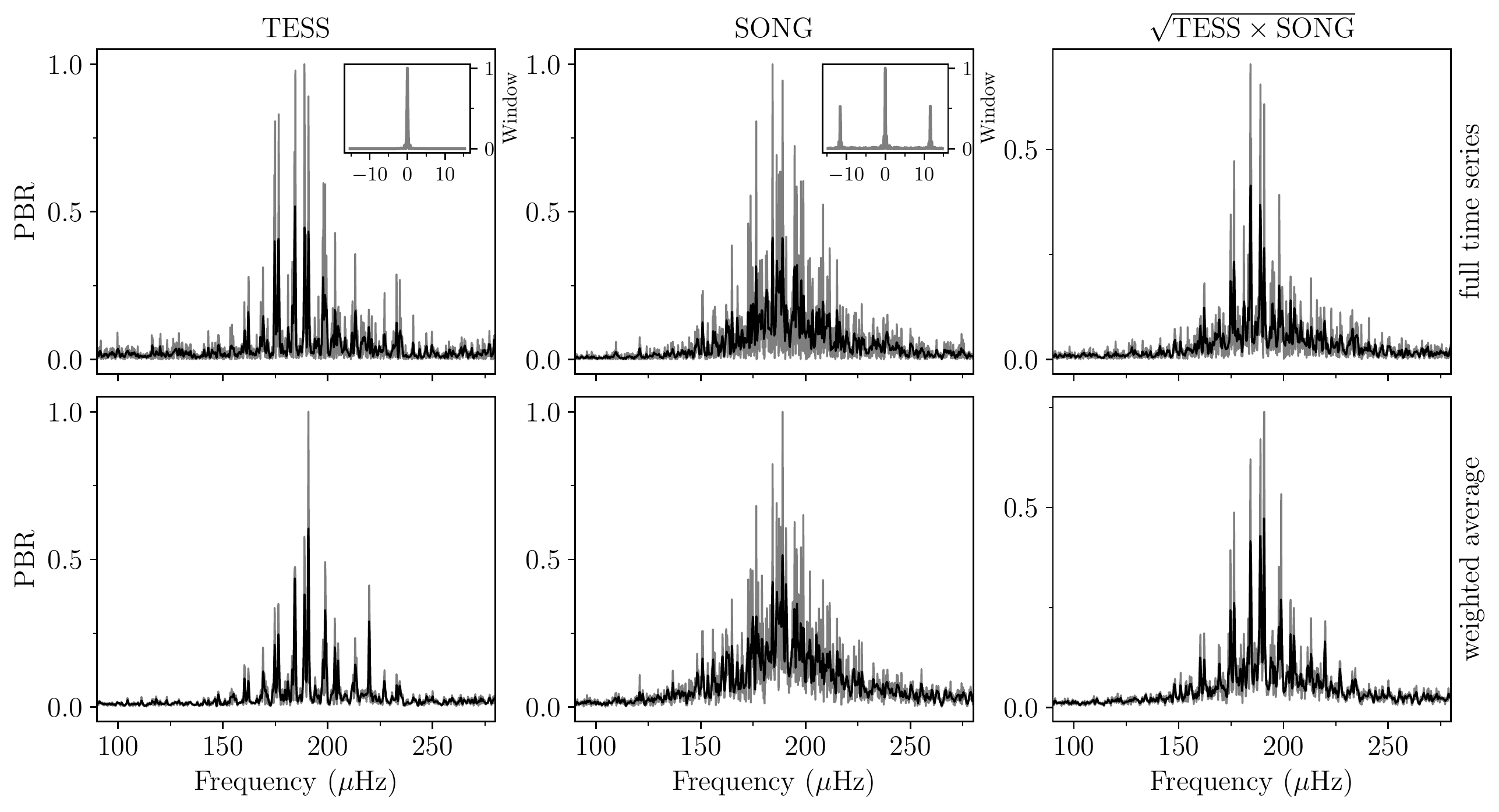}
    \caption{{\bf TESS and SONG background-corrected, normalised power spectra.} The power-to-background ratio (PBR) power spectra from TESS (left column) and SONG (middle) are shown in grey, and a smoothed (box-kernel) version is shown in black. The bottom row shows the weighted average power spectra, where we have taken the power spectrum for each sector or each campaign, divided by the fitted background, and created a weighted average. In the top row we have simply removed (divided) the background from the power spectra resulting from combining all sectors or all campaigns into the time series. The right column shows the product power spectra created from the full time series (top) and the weighted average power spectra (bottom). The insets in the top-left and top-middle panels show the spectral window from the full time series from \tess and SONG, respectively.}
    \label{fig:psd}
\end{figure*}

As we have multiple campaigns of SONG data, as well as multiple sectors of TESS observations, there are multiple ways of combining the data. We used a number of different power spectra in the vetting of our analysis of the seismic content, especially for oscillation mode identification. \fref{fig:psd} shows for SONG and TESS the power spectra from either the full time series or the weighted averaged power spectra from yearly (SONG) or sector-wise power spectra, with the weighting given by the inverse variance ($1/\sigma^2$) of the median spectral white noise level from $3,800$~$\mu$Hz to $3,900$~$\mu$Hz -- the average spectra are useful for the detection of oscillations from their higher signal-to-noise ratio, while the full spectra have superior frequency resolution (see \fref{fig:psd}).

Finally, we produced product spectra between the SONG and TESS data, where we used the product of the power spectra from the full time series in our frequency extraction (\sref{sec:freq_ex}). The benefit of the product spectra is the significantly reduced aliasing effect of the spectral window of SONG data. To combine the spectra we first fit and remove their granulation background signal, and then normalise to the peak envelope power at \numax. We adopt a common resolution of ${\sim}0.16\rm\,\mu Hz$ for the two spectra, given by the effective observing time for the SONG observation as found from the integral of the spectral window. 

To account for the granulation background, we modelled the power spectra as

\begin{equation}
    \mathcal{P}(\nu) = \eta^2(\nu) \left( L(\nu) + N (\nu) \right) + W \, ,
    \label{eq:power_lorentz}
\end{equation}
where $L$ is the Lorentzian accounting for the oscillation power excess, centred on $\nu_\mathrm{max}$ and with a width $\Gamma$ and amplitude $A$. Traditionally, the power excess is modelled by a Gaussian function, but high S/N spectra suggest the envelope is better approximated by a Lorentzian (Lund et al, in prep.). $W$ is the contribution from white noise, and $\eta^2(\nu)$ is the apodisation of the signal power at frequency $\nu$ from the 2 minute sampling of the temporal signal \citep[e.g.][]{Chaplin2011,Kallinger2014,Lund2017} given as $\eta^2(\nu)=\sin^2(x)/x^2$ with $x = \pi \nu \delta t$, where $\delta t$ is the integration time here given as the TESS 2-minute cadence. The granulation background is modelled with a power-law function as \citep{Harvey1993,Aigrain2004,Michel2009,Karoff2012,Lund2017}
\begin{equation}
    N (\nu) = \sum^{2}_{i=1} \frac{\xi_i \sigma^{2}_i \tau_i}{1 + (2 \pi \nu \tau_i)^{\alpha_i}} \, ,
    \label{eq:granu}
\end{equation}
corresponding to an exponentially decaying autocorrelation function, with a power of the temporal decay rate $-2/\alpha_i$. $\tau_i$ and $\sigma_i$ are respectively the characteristic timescale and the root-mean-square (rms) in the time domain of the $i$th background component, and $\xi_i$ gives the corresponding normalisation constant to ensure that Parseval's theorem is followed \citep[][]{Michel2009,Karoff2013}.

We employed a similar strategy to remove the background in the SONG data. However, as the granulation background is much less visible in RV, we only included one term in the granulation background, meaning $i=1$ in \eref{eq:granu}. We furthermore only included frequencies above $35$~$\mu$Hz, since we have essentially introduced a high-pass filter by removing the nightly offsets in the SONG RVs. 

In \tref{tab:seis_model} we give the resulting values for $\nu_\mathrm{max}$ from SONG and TESS, but caution that the uncertainties provided there are only internal. $\nu_\mathrm{max}$ as measured in this way from, for instance, the different SONG campaigns varies beyond the quoted error, which could be caused by variations in mode excitation and influenced by the correction for the inherently weak granulation background in RV. The tabulated value is chiefly governed by the 2014 data ($\sim$$191$~$\mu$Hz).

\subsection{Frequency extraction}\label{sec:freq_ex}

The extraction of individual oscillation mode frequencies (\ie peak bagging) was performed on the square-root of the full time series product power spectrum described above. In doing so we assume that aspects such as differences in mode asymmetries and relative amplitudes will not significantly impact the mode frequency determination, and that the modes are still well described by simple Lorentzian profiles -- given the relatively low spectral resolution of ${\sim}0.16\rm\,\mu Hz$ as compared to the expected mode line width for a star like \gamcep of $\rm {\sim}0.1-0.3\,\mu Hz$ \citep[\eg][]{Baudin2011,Corsaro2012,Handberg2017,legacy} this is a fair assumption. The peak bagging procedure overall followed that outlined in the LEGACY project by \citet{legacy}, but with a modification of the relation between mode amplitude and height as given by \citet{Fletcher2006} (their Eq. A4) \citep[see also][]{Chaplin2008} from the fact that the modes are not fully resolved. We also decoupled the determination of mode line widths and amplitudes for the mixed dipole modes from those of the radial modes.
Finally, we note that in constructing the square-root product spectrum one will inevitably alter the noise statistic from the usual $\chi^2$ $2$-d.o.f. distribution of the individual power density spectra (PDS). We find a noise distribution closer to that of a $\chi^2$ $12$-d.o.f. distribution, which we confirm is as expected from simulated data. In the peak bagging we try fitting both assuming the standard $\chi^2$ $2$ d.o.f. noise and adopting a modified likelihood corresponding to a $\chi^2$ $12$ d.o.f. noise \citep[see \eg][]{Appourchaux2003,phdthesis_lund}. We find no significant differences in the extracted mode frequencies from the two approaches.

\begin{figure*}
    \centering
    \includegraphics[width=\textwidth]{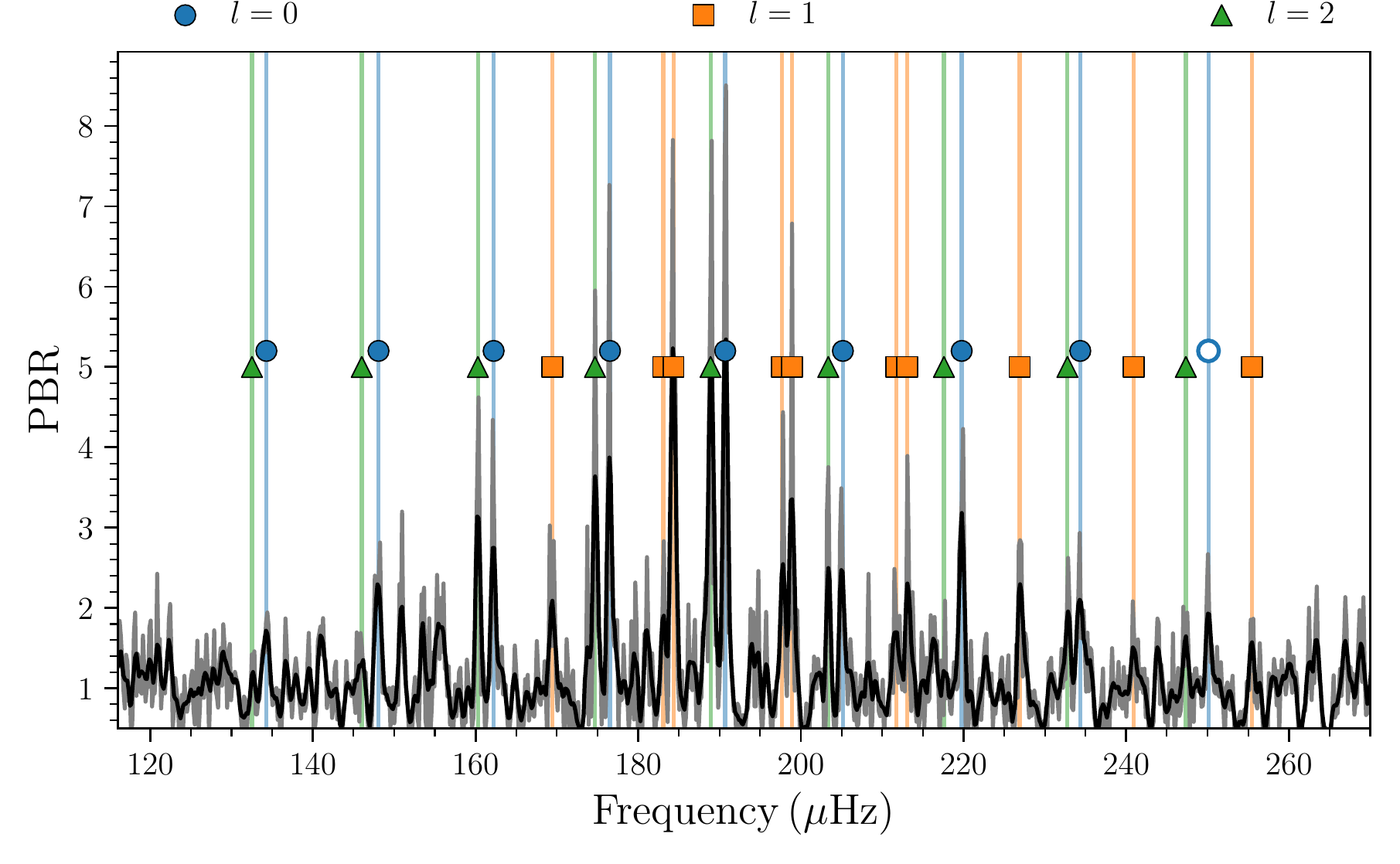}
    \caption{{\bf Power-to-background ratio} (PBR) of the product power spectrum ($\sqrt{\mathrm{SONG} \times \mathrm{TESS}}$) used for the peak bagging. The black spectrum is a smoothed version of the corresponding grey spectrum. The markers and vertical lines indicate the frequencies of the extracted oscillation modes.}
    \label{fig:new_psd}
\end{figure*}

\begin{figure}
    \centering
    \includegraphics[width=\columnwidth]{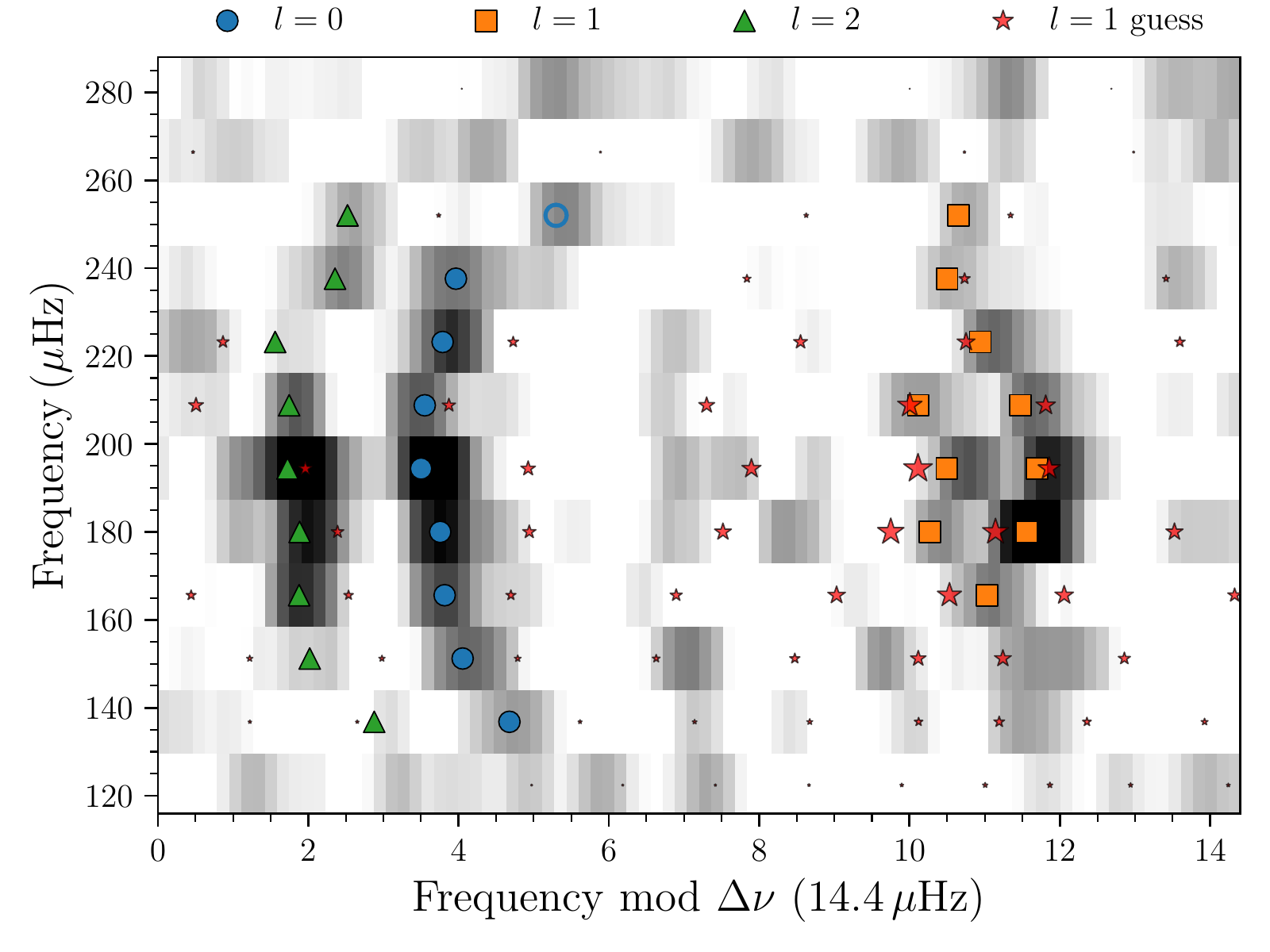}
    \caption{{\bf \'{E}chelle diagram} of the smoothed (black) spectrum in \fref{fig:new_psd}. The markers indicate the extracted modes. The red stars give the frequencies estimated from the asymptotic mixed-mode relation, with their size showing the expected relative amplitude of the modes. The empty marker indicate the fitted $l=0$ that we suspect is not a bona fide mode.}
    \label{fig:ech}
\end{figure}

The identification of oscillation modes was straightforward for the radial ($l=0$) and quadrupole ($l=2$) modes from the \'{e}chelle diagram and a value for \dnu. 
For the mixed dipole ($l=1$) modes we first manually identified the prominent peaks in the power spectrum that could not be associated with $l=0,2$, these are shown in the power spectrum in \fref{fig:new_psd} and \'{e}chelle diagram in \fref{fig:ech}. To evaluate the likelihood of a given peak actually being a mixed mode, we tried to match the asymptotic relation for mixed modes by  \citet{Shibahashi1979}, including the curvature of the variation in the large separation to be as in \citet{App2020} \citep[see also, e.g.,][]{Mosser2013}. With values for the coupling strength ($q$), \dnu, the phase factors $\epsilon_{\rm g}$ and $\epsilon_{\rm p}$, the strength of the \dnu variation/curvature ($\alpha$), $\delta\nu_{01}$ from the asymptotic p-mode relation, and the asymptotic period spacing ($\Delta\Pi_1$) one can solve for the frequencies of the mixed modes. The parameters pertaining to the p-modes ($\dnu, \epsilon_{\rm p}, \alpha$) are obtained from a fit to the identified radial modes following \citet{legacy}. A good first guess on $\Delta\Pi_1$ can be obtained from its proportionality to \dnu before the star enters the red clump phase \citep{Bedding2011,Mosser2014}, leading to a value of $\Delta\Pi_1\sim 84$ s. An estimate of the coupling factor of $q\sim 0.15\pm0.03$ is obtained from the results of \citet{Mosser2017}. From \citet{Mosser2018} we find $\epsilon_{\rm g}\sim 0.25\pm0.05$, and lastly we estimate $\delta\nu_{01}\sim -0.3\pm0.1\rm \,\mu Hz$ from \citet{Huber2010}.

To further aid the identification of potential dipole mixed modes we compute a proxy for the expected relative amplitudes of the asymptotically derived modes. This follows from the prescription of \citet{Benomar2014}, who finds that the dipole amplitude ($A_1$) can be found in units of the radial mode amplitudes ($A_0$) as $A_1 \approx A_0 V_1 (I_0 / I_1) \sqrt{\Gamma_0/\Gamma_1}$. Here $I_0 / I_1$ denotes the ratio of mode inertias, $\Gamma_0/\Gamma_1$ gives the ratio in mode line-widths, while $V_1$ gives the relative mode visibility \citep[][]{Ballot2011}, where we adopt $V_1^2=1.5$ \citep[see][]{Li2020}. This relation can be rewritten in terms of the stretch function $\zeta$ introduced by \citet{Mosser2014}, quantifying the degree of mode trapping \citep[see also][]{Jiang2014,Mosser2015,Hekker2017}, as $A_1 = A_0 V_1 \sqrt{1 - \zeta}$. We model $A_0(\nu)$ as a Gaussian with a full width half maximum of $\rm 60\,\mu Hz$

For the peak bagging we took the conservative approach and only selected potential modes where we could find an expected mode from the asymptotic relation in near proximity, and where the amplitude followed the expected pattern. We note that the spectrum contains a number of additional peaks that we could not easily associate with mixed modes. 
\fref{fig:ech} shows the frequencies and expected relative amplitudes for the mixed modes obtained from the asymptotic relation, in addition to the frequencies extracted from the peak bagging. 
The extracted frequencies are provided in \tref{tab:freqs}. We note that while the highest frequency $l=0$ mode corresponds to a significant excess power (\fref{fig:new_psd} and \fref{fig:ech}), we suspect that this is not related to an actual $l=0$ mode from the departure of the otherwise smooth ridge in the échelle diagram. We found, however, that this mode had a negligible impact on the seismic modelling.

\section{Seismic modelling}\label{sec:model}

To derive physical properties for \gamcepa, we compared our measured frequencies to those calculated from stellar modelling. We modelled the extracted frequencies, spectroscopic parameters, and the calculated luminosity using the BAyesian STellar Algorithm (BASTA) modelling pipeline as described below.

\subsection{BASTA}

As mentioned in \sref{sec:freq_ex}, mode identification for the mixed dipole modes is not as straightforward as for the $l=0,2$ modes. Therefore, we extracted and modelled our frequencies, namely the dipole modes, in an iterative manner. For this we used BASTA \citep[][]{BASTA2015,BASTA2022}.

BASTA fits a star using a grid of stellar models calculated from the Garching Stellar Evolution Code \citep[GARSTEC;][]{Weiss2008} along with Bayesian statistics in search of the best-fitting parameters. To accurately reflect the expected distribution of stars with lower mass stars being more abundant, the Salpeter initial mass function \citep{Salpeter1955} is applied. Frequency fitting in BASTA is done using the Aarhus adiabatic oscillation package \citep[ADIPLS;][]{Christensen2008} with the inclusion of a two-term surface correction as given by \citet{Ball2014}.

We ran BASTA with the observed frequencies listed in \tref{tab:freqs}, using the spectroscopic parameters (\teff, $\log g$, and $\rm [Fe/H]$) from \tref{tab:star} along with the \gaia luminosity from \tref{tab:star} as constraints. For the final iteration, we ran BASTA both with and without an additional constraint from our calculated luminosity. The two runs were in agreement, but the luminosity constraint naturally provided smaller uncertainties on the resulting parameters. We therefore adopt the values from this run as our final results.

\begin{figure*}
        \centering
        \begin{multicols}{2}
            {\includegraphics[width=\linewidth]{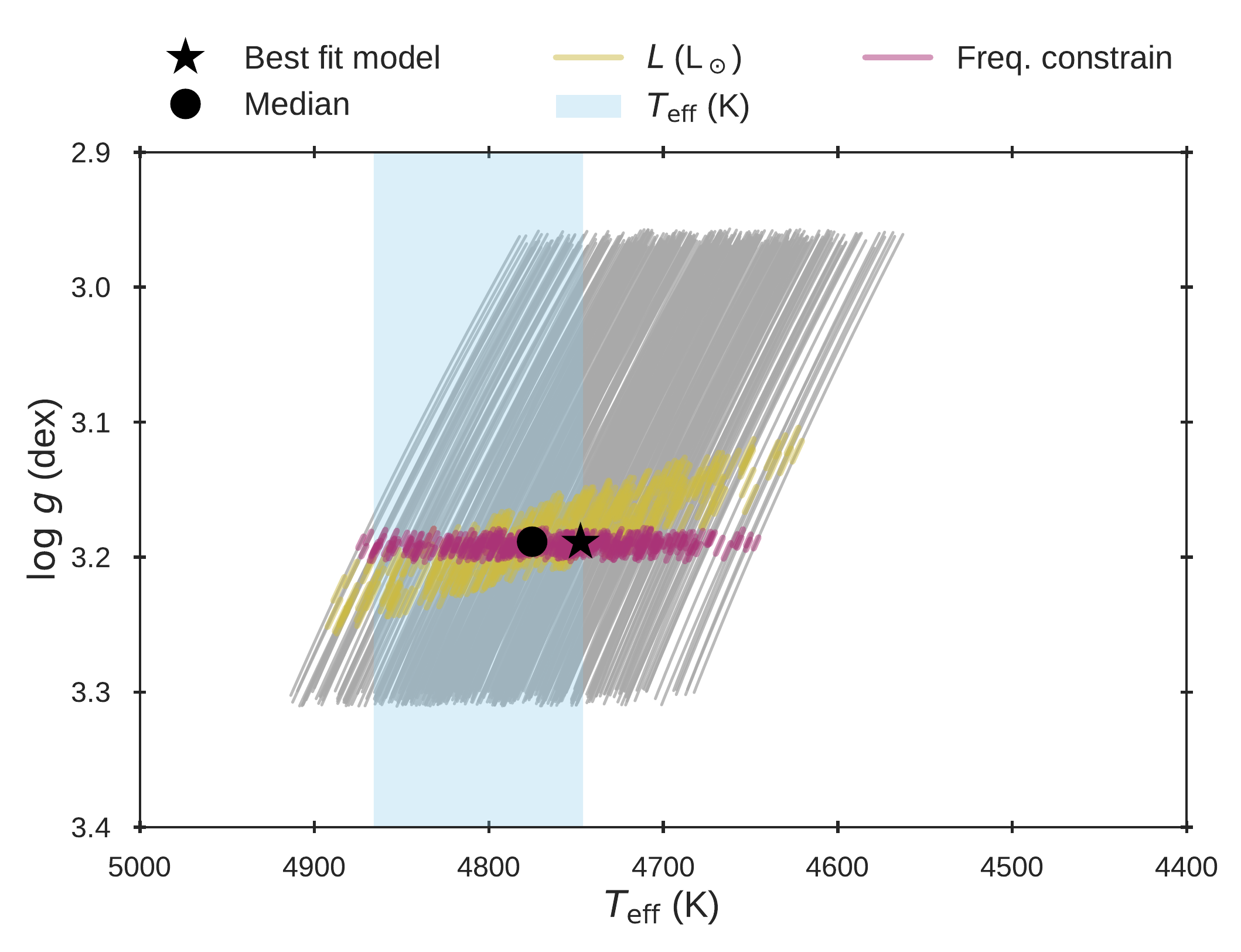}}\par 
           {\includegraphics[width=\linewidth]{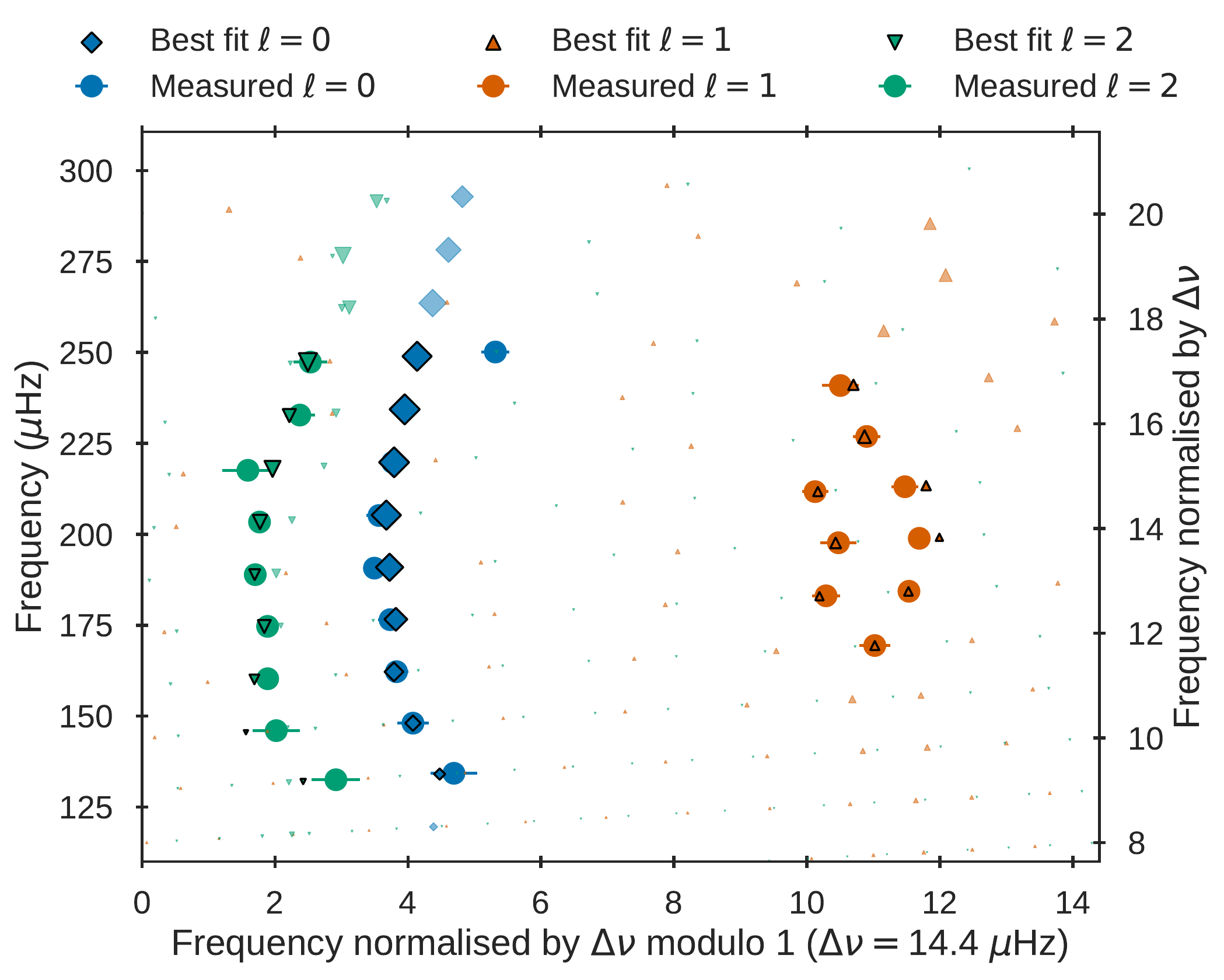}}\par 
               \end{multicols}
    \caption{{\bf BASTA Kiel and \'{e}chelle diagrams. } Left: Kiel diagram showing the model grid in grey, with the best fitting and median values denoted by a star and circle, respectively. The constraints applied for the effective temperature, luminosity, and frequencies are shown as the blue, yellow, and magenta shaded ares, respectively.  Right: \'{E}chelle diagram showing our measured $l=0$ (blue), $l=1$ (orange), and $l=2$ (green) frequencies as circles with error bars compared to our model frequencies from BASTA, shown with black outlines. Transparent markers with no black outline denote frequencies we have not detected. As in \fref{fig:ech}, marker sizes show the expected relative amplitude of the modes.}\label{fig:ech_basta}
\end{figure*}

\begin{figure}
    \centering
    \includegraphics[width=\columnwidth]{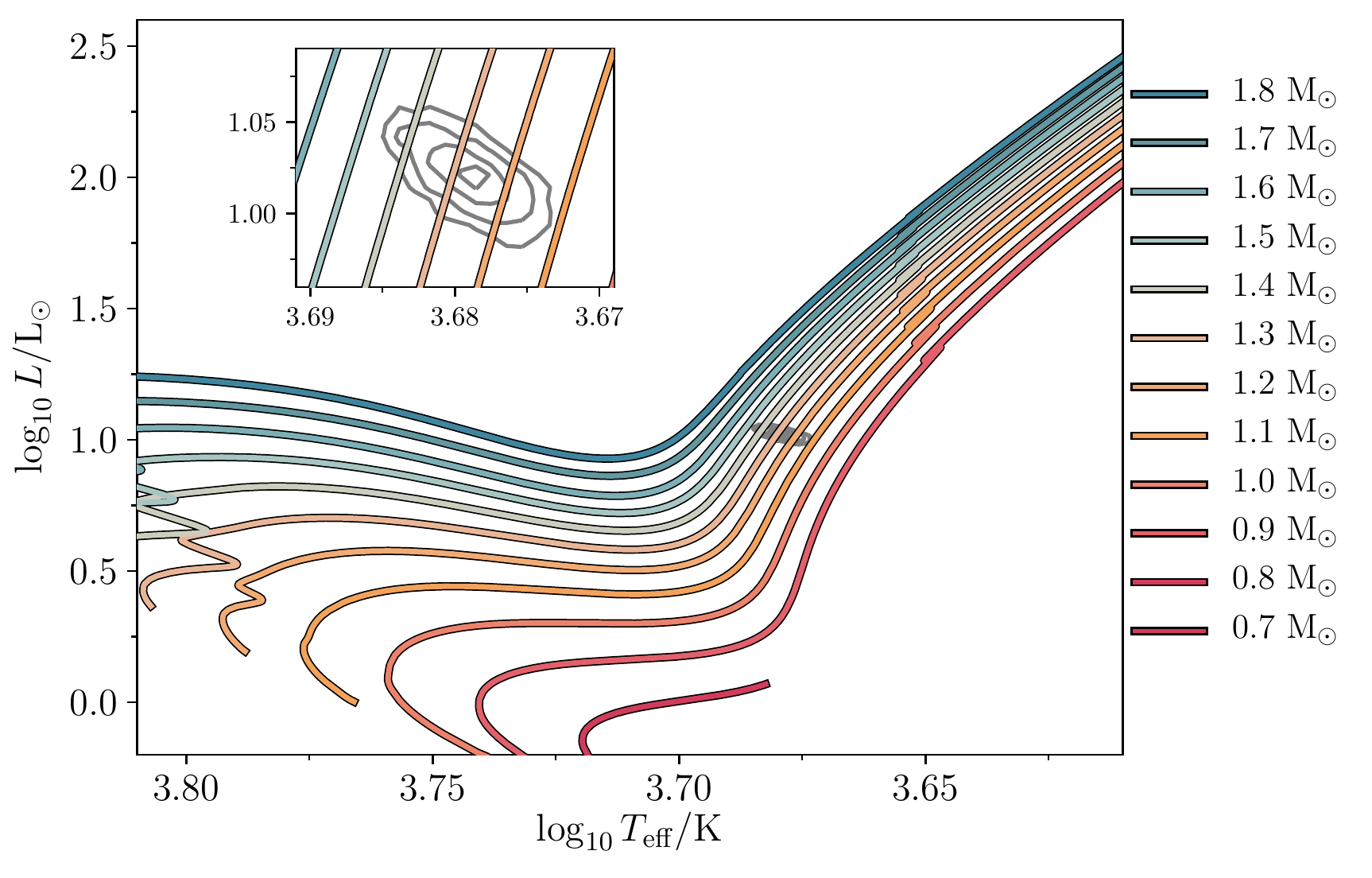}
    \caption{Hertzsprung-Russell diagram with the GARSTEC tracks used for the fit with BASTA. The tracks are spaced by $0.1$~M$_\odot$ and all have a metallicity of $[\rm Fe/H]=0.2$~dex. The location of \gamcepa is shown with contours created from the BASTA \teff and $R_\star$ in \tref{tab:seis_model}. The inset shows a close-up around \gamcepa.}
    \label{fig:hr_diag}
\end{figure}

The resulting fit to the models and their frequencies from BASTA is shown in \fref{fig:ech_basta}, and we show the distributions for the physical properties and global seismic parameters in the correlation plot in \fref{fig:corner}. We find a mass of \finalmass, a radius of \finalradius, and an age of \finalage, which we have tabulated in \tref{tab:seis_model} along with the spectroscopic parameters (mainly reflecting the input). For the global seismic parameter BASTA finds \numax$=$\finalnumax and \dnu$=$\finaldnu. As expected, the results from BASTA suggests that \gamcepa is an RGB star as shown in the Hertzsprung-Russell diagram in \fref{fig:hr_diag}.

\subsection{Interferometry}
We can obtain an independent measurement of the stellar radius by combining the angular diameters from interferometry ($\theta_{\rm IF}$) with the distance from \gaia as\begin{equation}\label{eq:inter_r}
    \frac{R_{\mathrm{IF}}}{\mathrm{R_{\odot}}} = \frac{D_{\emph{Gaia}}\times\mathrm{AU}}{2\mathrm{R_{\odot}}}\theta_{\mathrm{IF}}\,\, ,
\end{equation}
with $\theta_{\mathrm{IF}}$ in arcseconds, the distance $D_{\emph{Gaia}}$ in parsec (\tref{tab:star}), and with $\mathrm{AU}$ giving the astronomical unit\footnote{We adopt 1 AU $=149.5978707 \times 10^9$~m (IAU 2012 Resolution B2) and $\mathrm{R_{\odot}}=6.957\times 10^8$~m \citep[IAU 2015 Resolution B3;][]{IAUsolarR}.} \citep[see][]{Victor2012, keystone}.
In \tref{tab:inter} we list the different available values from the literature together with the resulting stellar radius when adopting the \gaia DR3 value for the distance by \citet{BailerJones2021}.

We note that the \gaia data provides a high re-normalised unit weight error  \citep{ruwe2018} parameter for \gamcepa, which, at a value of $3.212,$ is significantly higher than the suggested threshold of $1.4$. This suggests that the \gaia distances might be affected by a sub-optimal astrometric solution, likely because of the orbital motion induced by the binary companion.

\begin{table}
    \centering
    \caption{{\bf Interferometry.}  }
    \begin{threeparttable}
    \begin{tabular}{c c c c}
\toprule 
Source & Instrument & $\theta$ & Radius \\ 
& & (mas)  & ($\rm R_{\odot}$) \\ 
\midrule
\citet{Baines2018} & NPOI & $3.254\pm0.020$ & $4.81\pm0.03$ \\ 
\citet{Hutter2016} & NPOI & $3.329\pm0.042$ & $4.92\pm0.06$ \\ 
\citet{Baines2009} & CHARA & $3.302\pm0.029$ & $4.88\pm0.04$ \\ 
\citet{Nordgren1999} & NPOI & $3.24\pm0.03$ & $4.79\pm0.05$ \\ 
\bottomrule
    \end{tabular}
    \begin{tablenotes}
    \item Angular diameters obtained from interferometry and the associated stellar radius (Eq.~\ref{eq:inter_r}) when adopting the \emph{Gaia} DR3 distance from \citet{BailerJones2021} (see \tref{tab:star}).
    \item NPOI: Navy Prototype Optical Interferometer
    \item CHARA: Center for High Angular Resolution Astronomy
\end{tablenotes}

    \end{threeparttable}
    \label{tab:inter}
\end{table}

\section{Discussion}\label{sec:disc}

\subsection{Seismology and physical properties}

As is characteristic for subgiants and RGB stars, the power spectrum of \gamcepa shows mixed modes, which appears when the central region of the star starts to contract, increasing the density of the core. This increases the frequencies of the g-modes, and they will start to couple strongly to the p-modes. Although this complicates the frequency extraction, as the modes no longer follow the asymptotic relation for pure acoustic modes, the mixed modes also allow us to probe the internal properties as they provide useful information about the core.

We find an excellent agreement between the observed dipole modes and those calculated by ADIPLS when fitting with BASTA (\fref{fig:ech_basta}), despite the strong mixing. Furthermore, the radial and quadrupole modes also agree very well with the model, with the exception of the highest order ($n=17$) observed $l=0$ mode (\sref{sec:freq_ex}). The excellent agreement allows us to place rather tight constraints on parameters such as the mass (${\sim}5\%$), the radius (${\sim}2\%$), and the age (${\sim}14\%$). 

\subsection{The binary and planetary system}

In an attempt to refine the orbital parameters for both the binary and the planetary orbit, we modelled the orbit in a fit using our RVs from the different SONG campaigns. When we initially started modelling the orbit we noticed a prominent signal of around 90~d after having subtracted the RV signals of the companion star, $\gamma$~CepB, and planet, $\gamma$~CepAb, as reported in \citet{Hatzes2003}. A number of instrumental effects could introduce a signal of a few metres per second. Temperature and pressure changes being obvious candidates are however unlikely given the use of an Iodine cell for wavelength calibration. Long-term trends are known to be present for a number of stars observed with SONG as mentioned in \citet{Arentoft2019}. These show a velocity change of typically non-sinusoidal nature with a periodicity of 1~yr. The exact morphology and amplitude is strongly dependent on the sky position of the objects, namely the proximity to the ecliptic. 

The variations for \gamcep, however, looked more sinusoidal, which might be because we have mainly been sampling the same phase of the 1~yr signal, namely in the (northern hemisphere) autumn. The 90~d signal is thus an artefact of the known 1~yr signal. Therefore, we included an additional term in our model with a period fixed to 365.25~d, but allowing the phase and amplitude to vary. As we have data from the Hertzsprung SONG node on Tenerife that were obtained with two different iodine cells (only in 2017), we include two separate RV offsets, $\Gamma$, for these, and we naturally apply a third offset for the data acquired with the Chinese node at Delingha.

As the oscillations are clearly seen in the SONG data, we modelled these using Gaussian process (GP) regression utilising the library \texttt{celerite} \citep{celerite}. Our GP kernel is composed of two \texttt{SHOTerm} components in \texttt{celerite}, which are stochastically driven, damped harmonic oscillators. The \texttt{SHOTerm} is characterised by three hyperparameters governing the behaviour of the kernel; the power, $S_0$, the quality factor or line width, $Q$, and the undamped angular frequency, $\omega_0$, which here is directly related to $\nu_{\rm max}$. The first term in our kernel is meant to capture the oscillation signals, and the second term is designed to capture any longer term variability, like granulation, although we do not expect that to be prominent in the RVs. 

\begin{figure}
    \centering
    \includegraphics[width=\columnwidth]{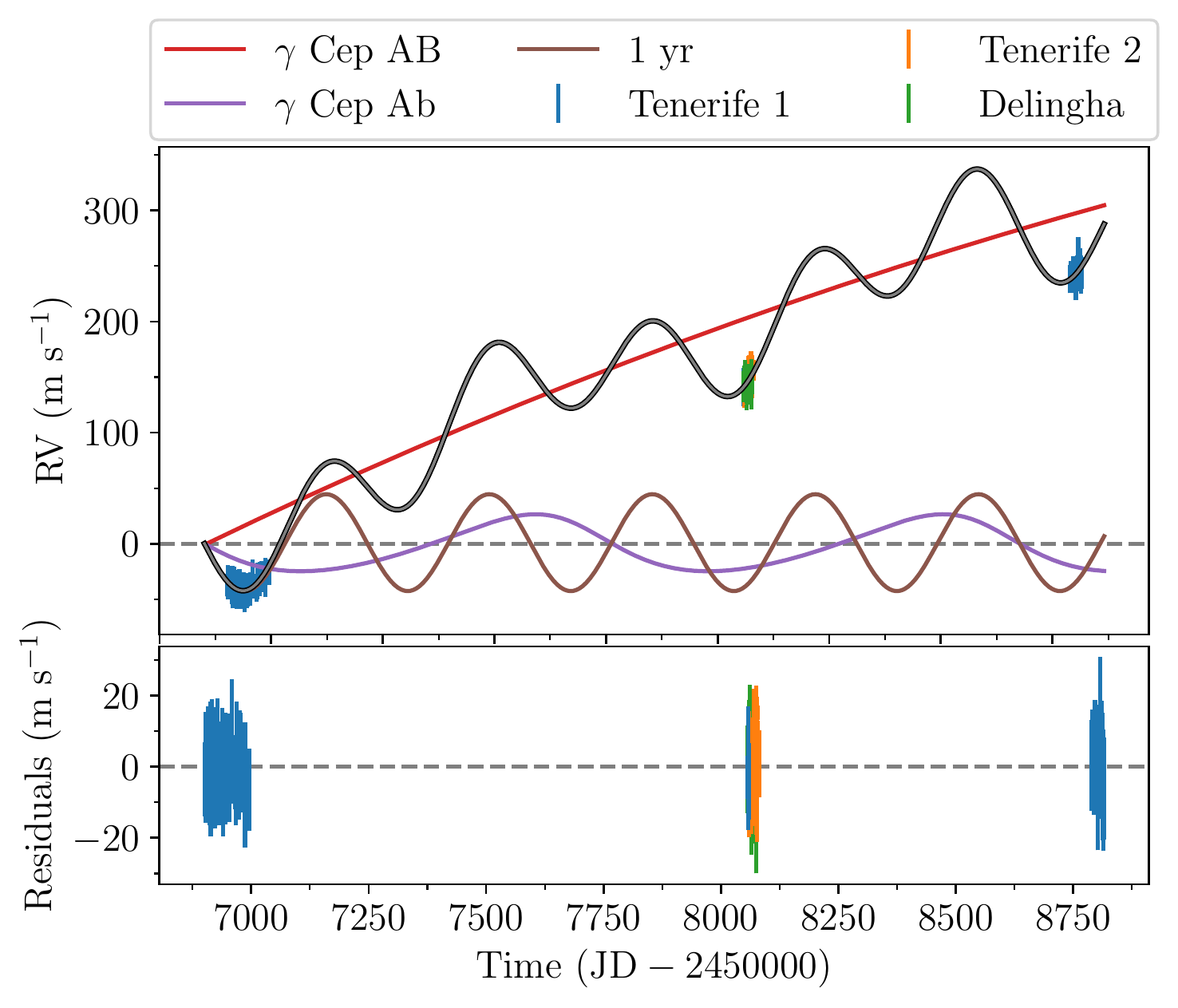}
    \caption{{\bf Orbital motion.} {\it Top:} Two {\it Keplerian} orbits from \gamcepb and \planet shown in red and purple, respectively, and the 1~yr signal shown as the brown curve. The sum of the signals is shown as the grey model. Here they have all been shifted to start at 0.0 to make it easier to compare them. The Tenerife data are shown in blue and orange, and the Delingha data are shown in green. {\it Bottom:} Residuals after subtracting the model. }
    \label{fig:orbit}
\end{figure}

\begin{figure}
    \centering
    \includegraphics[width=\columnwidth]{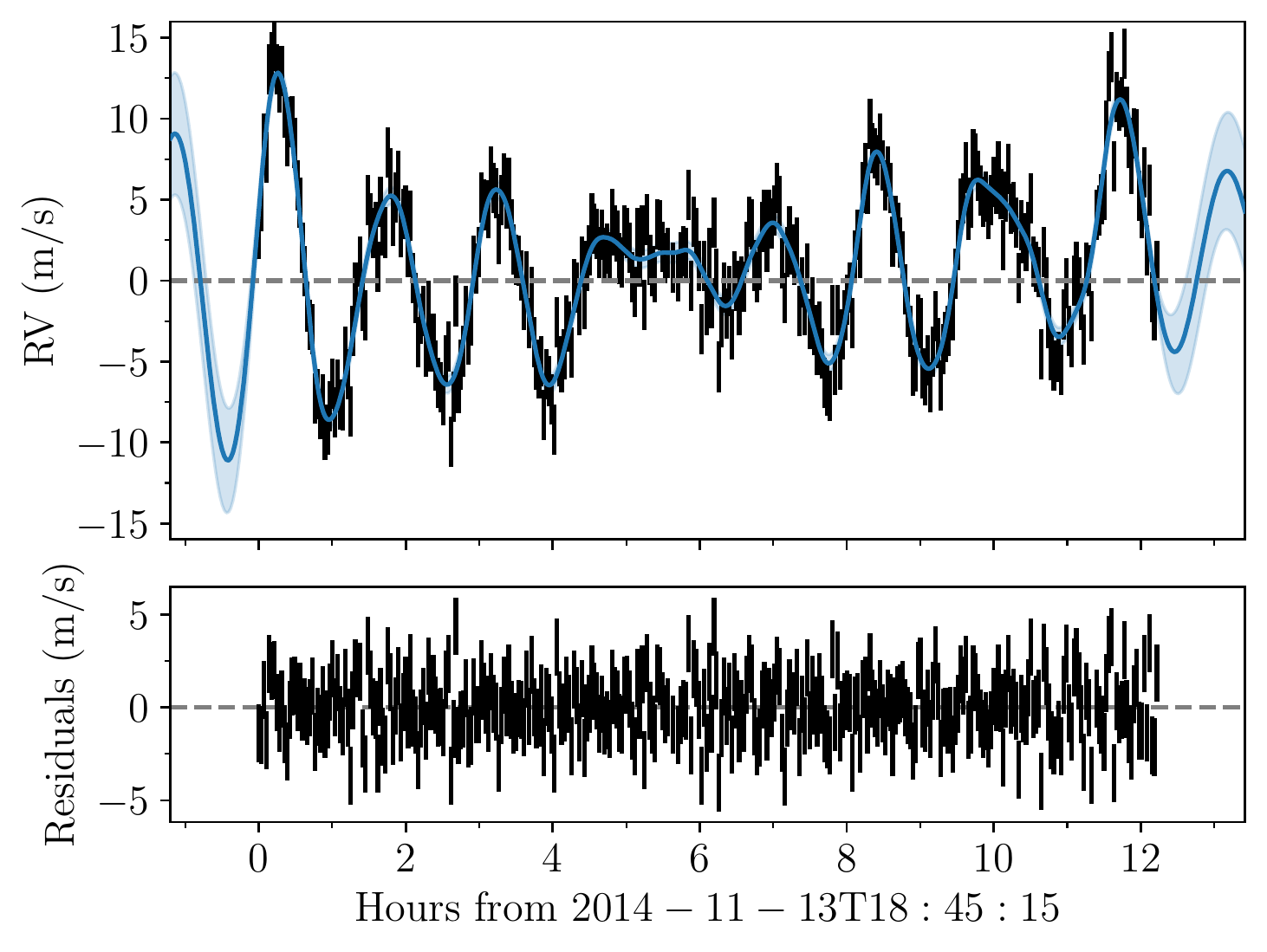}
    \caption{{\bf GP detrended SONG data.} {\it Top:} Snippet of SONG data from the 2014 campaign (same snippet as in \fref{fig:song_rv}) after subtracting the signals shown in \fref{fig:orbit}. The blue line is the GP model included in the fit, with the transparent band showing the 1$\sigma$ confidence interval. {\it Bottom:} Residuals after subtracting the GP model.}
    \label{fig:gp_song}
\end{figure}

Following \citet{Pereira2019} we fix the quality factor in our granulation \texttt{SHOTerm} term (SHO2) to $Q_2=1/\sqrt{2}$, and we further found it necessary to fix the amplitude for this term, $S_2$ -- the amplitude is weakly constrained by, generally, being lower than that of the oscillations and only dominate the data at the longest timescales (lowest frequencies; see \fref{fig:gp_song}). This term had a tendency to pick-up on the 1~yr signal, resulting in a significantly poorer fit when subtracting the orbital motion from \gamcepb and \planet as well as the resulting amplitude and phase for the 1~yr signal (as done for a fit with a fixed value for $S_2$ in \fref{fig:orbit}). The amplitude, $S_2$, thus seemed to over-compensate as the overall residuals were seemingly identical between having this parameter fixed or free to vary (as seen in \fref{fig:gp_song} again for a fixed value for $S_2$). Finally, we also include a white noise or jitter term, where we fix the value for the hyperparameter, $\sigma$ (see \tref{tab:orbit}).

In our fit we used priors from \citet[][their $\chi^2=1.44$ solution]{Huang2022} for the binary and planetary orbit. These priors along with all parameters are summarised in \tref{tab:orbit}. We sampled the posterior distribution through a Markov chain Monte Carlo (MCMC) analysis using \texttt{emcee} \citep{emcee} with 100 walkers. We ran the MCMC until convergence, which we assessed through the rank normalised $\hat{R}$ diagnostic test \citep{Vehtari2019} as implemented in \texttt{ArviZ} \citep{arviz}.

\begin{table*}
    \centering
    \caption{{\bf Orbital parameters and hyperparameters.} }
    \begin{threeparttable}

    \begin{tabular}{c c c c}
        \toprule
        Parameter & Description & Prior & Value \\
        \midrule
        $K_{\rm AB}$ (m~s$^{-1}$) & Binary RV semi-amplitude & $\mathcal{N}(1699.94,3.32)$ & $1711\pm3$  \\
        $P_{\rm AB}$ (d) & Binary orbital period & $\mathcal{N}(20731.68,58.36)$ & $21170_{-58}^{+48}$ \\
        $T_{\rm 0,AB}$ ($\rm JD-2400000$) & Time of inferior conjunction for binary orbit & $\mathcal{N}(48435.04,0.67)$ & $48435.2_{-0.7}^{+0.6}$ \\
        $e_{\rm AB}$ & Eccentricity of binary orbit & $\mathcal{N}(0.3605,0.0026)$ & $0.333 \pm 0.002$ \\
        $\omega_{\rm AB}$ ($^{\circ}$)& Argument of periastron of binary orbit & $\mathcal{N}(158.90,0.2)$ & $157.07^{+0.18}_{-0.15}$  \\
        $K_{\rm Ab}$ (m~s$^{-1}$) & Planetary RV semi-amplitude & $\mathcal{N}(26.40,1.30)$ & $25.6 \pm 1.3$ \\
        $P_{\rm Ab}$ (d) & Planetary orbital period & $\mathcal{N}(901.46,2.84)$ & $913 \pm 3$ \\
        $T_{\rm 0,Ab}$ ($\rm JD-2400000$) & Time of inferior conjunction for planetary orbit & $\mathcal{N}(53107.63,28.19)$ & $53117^{+18}_{-16}$ \\
        $e_{\rm Ab}$ & Eccentricity of planetary orbit & $\mathcal{N}(0.0856,0.075)$ & $0.15^{+0.07}_{-0.05}$ \\
        $\omega_{\rm Ab}$ ($^{\circ}$)& Argument of periastron for planetary orbit & $\mathcal{N}(55.37,6.7)$ & $49^{+6}_{-7}$ \\
        \hline
        $P$ (d) & Period of 1~yr signal & $\mathcal{F}(365.25)$ & - \\
        $\log_{10} t$ ($\rm \log_{10} JD$) & Phase of 1~yr signal & $\mathcal{U}(3,7)$ & $5.0681\pm 0.0013$ \\
        $K$ (m~s$^{-1}$) & Amplitude of 1~yr signal & $\mathcal{U}(5,50)$ & $43.6157^{+0.0009}_{0.0011}$ \\
        \hline
        $\Gamma_{1}$ (m~s$^{-1}$) & Velocity offset Tenerife 1 & $\mathcal{U}(25059,33259)$ & $28708 \pm 6 $ \\
        $\Gamma_{2}$ (m~s$^{-1}$) & Velocity offset Tenerife 2 & $\mathcal{U}(25059,33259)$ & $28944^{+5}_{-6}$ \\
        $\Gamma_{3}$ (m~s$^{-1}$) & Velocity offset Delingha & $\mathcal{U}(-1300,0)$ & $-874 \pm 6$ \\

        \hline
        $\ln S_1$ ($\ln$~m$^{2}$~s$^{-2}$) & Amplitude & $\mathcal{U}(-5,5)$ & $-3.48 \pm 0.05$ \\
        $\ln \omega_1$ ($\ln$~d$^{-1}$) & Angular frequency & $\mathcal{U}(0,8)$ & $4.654 \pm 0.005$ \\
        $\ln Q_1$  & Quality factor & $\mathcal{U}(-2,5)$ & $1.91^{+0.06}_{-0.08}$ \\
        $\ln S_2$ ($\ln$~m$^{2}$~s$^{2}$) & Amplitude & $\mathcal{F}(-1.0)$ & - \\
        $\ln \omega_2$ ($\ln$~d$^{-1}$) & Angular frequency & $\mathcal{U}(0,8)$ & $2.48^{+0.07}_{-0.08}$ \\
        $\ln Q_2$  & Quality factor & $\mathcal{F}(\ln 1/\sqrt{2})$ & - \\
        $\ln \sigma$ ($\ln$~m~s$^{-1}$) & Jitter term & $\mathcal{F}(-9.4)$ & - \\
        \bottomrule
    \end{tabular}
    \begin{tablenotes}
    \item Orbital and GP hyperparameters in our MCMC analysis. The priors are denoted by $\mathcal{N}(\mu,\sigma)$ for a Gaussian prior with mean, $\mu$, and width, $\sigma$, $\mathcal{U}(a,b)$ for a uniform prior in the interval from $a$ to $b$, and $\mathcal{F}(c)$ for a parameter with a fixed value of $c$. 
    \end{tablenotes}
   \end{threeparttable}
    \label{tab:orbit}
\end{table*}

The results are given in \tref{tab:orbit}, and we show the resulting orbit in \fref{fig:orbit}. We furthermore show a close-up of a single night of observations in \fref{fig:gp_song}, where the oscillations and GP model are clearly seen. The root mean square (rms) for the residuals in \fref{fig:gp_song} comes out to around 1.56~m~s$^{-1}$, and calculating the rms for every night of observations we find a median nightly rms of around 1.36~m~s$^{-1}$. The typical rms reported in \citet[][Table 6]{Hatzes2003} is around 15~m~s$^{-1}$. If we do not remove the oscillations from our RVs, we get an rms of 4.75~m~s$^{-1}$.

In the context of exoplanet detection and characterisation it is important to be able to properly account for stellar oscillations and granulation, given that the limiting factor is not the precision of the (modern) spectrograph, but rather the intrinsic stellar signal. Here we are able to achieve this high precision because of the unique capabilities of SONG being dedicated to high-cadence monitoring. Sparsely sampled RVs, however, will be affected more strongly by this intrinsic stellar signal, although there are ways to partly circumvent the effects by optimising the integration time \citep{Chaplin2019}. This optimisation does, however, require knowledge of \numax.

\subsection{The masses}

From our mass measurement of \gamcepa in \tref{tab:seis_model} and the orbital elements in \tref{tab:orbit} we can estimate the mass of \planet using the mass function

\begin{equation}
    \frac{M^3_{\rm Ab} \sin^3 i_{\rm Ab}}{(M_{\rm Ab} + M_{\rm A})^2} = \frac{K^3_{\rm Ab} P_{\rm Ab} (1-e^2_{\rm Ab})^{3/2}}{2 \pi G} \, ,
    \label{eq:mass}
\end{equation}
with $G$ being the gravitational constant. We did this by drawing normally distributed values from our measurements in \tref{tab:seis_model} and \tref{tab:orbit}, while solving for $M_{\rm Ab}$ in \eref{eq:mass} in each of the 1,000 draws we did. From this we calculated the lower limit for the mass ($i_{\rm Ab}=90^\circ$), but we also expanded it to get an estimate for the actual mass. For the orbital inclination, $i_{\rm Ab}$, we took a conservative approach by drawing values uniformly between the boundaries from \citet{Reffert2011} who provided a lower limit of $i_{\rm Ab}=3.8^\circ$ and an upper limit of $i_{\rm Ab}=20.8^\circ$ (at 3$\sigma$ confidence). Similarly, we calculated the mass for \gamcepb using the orbital inclination from \citet{Neuhauser2007} of $i_{\rm AB} = 119.3 \pm 1.0 ^\circ$, where this time we drew normally distributed values. The results are given in \tref{tab:masses}. 

\begin{table*}
    \centering
    \caption{{\bf Masses of the bodies in \gamcep found in the literature. }}
    \begin{threeparttable}

    \begin{tabular}{c c c c c c c c c}
    \toprule

    Source & Method & $M_{\rm A}$ & Method & $M_{\rm B}$ & Method & $M_{\rm Ab}$ & $M_{\rm Ab}\sin i_{\rm Ab}$ \\
    & $M_{\rm A}$ & (M$_\odot$) & $M_{\rm B}$ & (M$_\odot$) & $M_{\rm Ab}$ & (M$_{\rm Jup}$) & (M$_{\rm Jup}$) \\
    \midrule
        \citet{Fuhrmann2004} & Spectroscopy & 1.59 & - & - & - & - & - \\ 
        \citet{Hatzes2003} & \citet{Fuhrmann2004} & $1.59 \pm 0.12$ & - & - & Derived & - & $1.7 \pm 0.4$ \\
        \citet{Torres2007} & Spec. \& phot. & $1.18 \pm 0.11$ & Derived & $0.362 \pm 0.022$ & - & - & $1.43 \pm 0.13$ \\
        \citet{Neuhauser2007} & Dynamical & $1.40 \pm 0.12$ & Dynamical & $0.409 \pm 0.018$ & - & - & $1.60 \pm 0.13$ \\
        \citet[][Table 5]{Mortier2013} & Spectroscopy & $1.26 \pm 0.14$ & - & - & - & - & - \\
        
        \citet{Stello2017} & Asteroseismology & $1.32 \pm 0.20$ & - & - & - & - & - \\
        \citet{Baines2018} & Interfero. (model) & $1.41 \pm 0.08$ & - & - & - & - & - \\ 
        \citet{Malla2020} & Asteroseismology & $1.32 \pm 0.12$ & - & - & - & - & - \\
        This work & Asteroseismology & $1.27^{+0.05}_{-0.07}$ & Derived & $0.328^{+0.009}_{-0.012}$ & Derived & $6.6^{+2.3}_{-2.8}$ & $1.41 \pm 0.08$ \\

         \bottomrule
    \end{tabular}
    \begin{tablenotes}
        \item Here we only report on values explicitly reported in any of the given papers. The method denotes the approach used to derive the quantity, the masses of $M_{\rm B}$ and $M_{\rm Ab}$ this are typically derived from the mass function, with the exception of dynamically determined masses. \citet{Hatzes2003} have used the mass estimate from \citet{Fuhrmann2004}, who states that typical errors for the mass are less than 10\% for the stars in that study.
    \end{tablenotes}
    \end{threeparttable}
    \label{tab:masses}
\end{table*}

We furthermore list mass estimates from the literature for all three bodies. These are determined from different approaches, where the mass of the primary has been determined spectroscopically as well as in combination with photometry, dynamically, and using asteroseismology. The secondary has typically been derived from the mass function assuming a mass for the primary, but has also been determined dynamically. The mass of \planet has exclusively been derived from the mass function.

\subsection{A long-period signal?}

\begin{figure}
    \centering
    \includegraphics[width=\columnwidth]{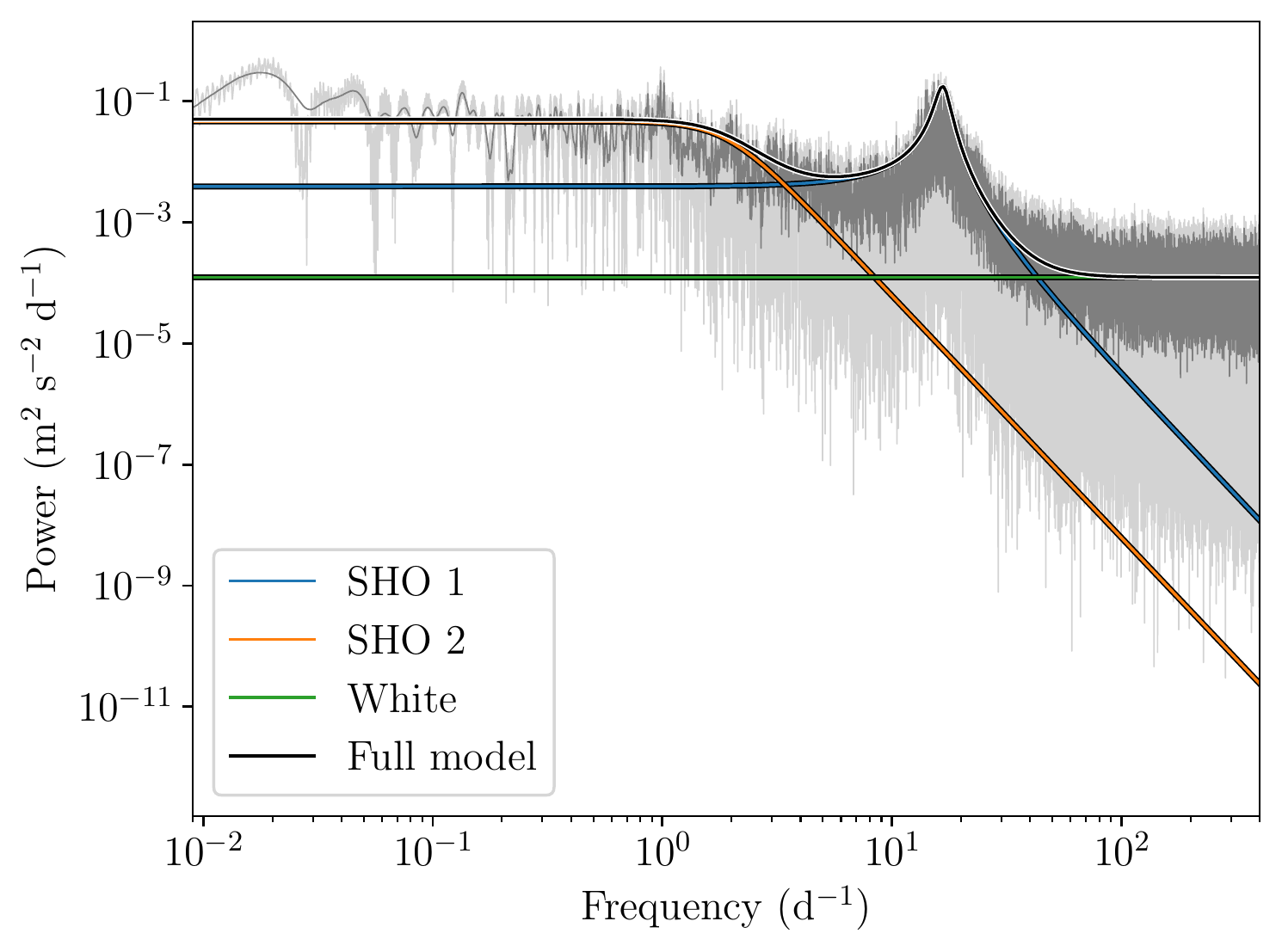}
    \caption{{\bf RV GP power spectrum.} The power spectrum of our GP model (black) is compared to the observed power spectrum (light grey) from our RVs after subtracting the orbital motion and 1~yr signal (\fref{fig:orbit}, bottom). In darker grey we show the observed power spectrum smoothed with a box kernel. The orange and blue curves show our harmonic GP terms, while the green line shows our white noise term. 
    }
    \label{fig:gp_power}
\end{figure}

In \fref{fig:gp_power} we show the power spectrum of both the observations and the GP model. Evidently, there is a hint of an excess at frequencies around $0.23$~$\mu$Hz ($0.02$~d$^{-1}$) corresponding to a period of 50~d. An additional source of variation in the system has been discussed previously, though at a significantly longer timescale. \citet{Hatzes2003} discuss the variation of the CaII $\lambda$8662 equivalent widths obtained by \citet{Walker1992}. As the variation is seen only in a specific time interval (1986.5-1992) over the course of the data acquisition up until that point, \citet{Hatzes2003} argue that it could be due to a period activity cycle of some 10-15~yr. While this could coincide with the SONG 2014, 2017, and 2019 campaigns, we find it unlikely that this could give rise to the signal we are seeing in the SONG data given that their period is 781~days and is only apparent in the CaII $\lambda$8662 equivalent width.

Another possibility is rotation. The studies by, for example, \citet{Garcia2014}, \citet{Ceillier2017}, and \citet{Santos2021} have investigated the rotational properties of stars observed by \kepler, including stars with solar-like oscillations. Their studies suggest typical rotation periods of 10-30~d for low-luminosity red giants (or subgiants), but with a good fraction of stars with longer periods, and 50~d is not uncommon. 

In the following we explore the implications IF the 50~d period signal is due to rotation. In that case, the period of the signal would be given by
\begin{equation}
    P_{\rm rot} = \frac{2 \pi R_{\rm A}}{v} \, ,
    \label{eq:prot}
\end{equation}
where $v$ is the rotation speed at the equator. If we then assume that $P_{\rm rot}=50\pm5$~d, and assume that the spots causing the rotational modulation are concentrated towards the equator, we can get the stellar inclination by substituting in \eref{eq:prot}: 

\begin{equation}
    \begin{split}
        i_{\rm A} &= \sin^{-1} \left ( \frac{v \sin i_{\rm A}}{v} \right ) \, , \\
            & = \sin^{-1} \left ( \frac{v \sin i_{\rm A}}{2 \pi R_{\rm A}/P_{\rm rot}} \right) \,.
    \end{split}
    \label{eq:inc}
\end{equation}
We followed the approach in \citet{Masuda2020} to account for the fact that $v$ and \vsini are not independent. Using $R_\star=4.67\pm0.15$~R$_\odot$ from BASTA (\tref{tab:seis_model}) and \vsini from \tref{tab:star} of $0.0 \pm 0.9$~km~s$^{-1}$ (truncated at $0.0$~km~s$^{-1}$), we get a value for the stellar inclination of $i_{\rm A}=13^{+9 \circ}_{-6}$, meaning that we are close to seeing the star pole-on.

Our measurement of a low \vsini is broadly consistent with previous studies, such as \citet[][\vsini$<0.3$~km~s$^{-1}$]{Walker1992} and \citet[][\vsini$=1.63\pm0.23$~km~s$^{-1}$]{Jofre2015}. We caution that at this level of projected rotation, disentangling rotation from macroturbulence in particular becomes challenging \citep[][]{Gray2005} but note that our value for $\zeta$ is in agreement with predictions from \citet[][]{Hekker2007} \citep[see also][]{Gray1989,Gray2005} covering the range $3.156-5.419$~km~s$^{-1}$ for the temperature and luminosity class of \gamcepa.

In any event, the imprint of rotation on the RV time series should be of low amplitude as the peak-to-peak variability from activity in RV can be approximated by the product of the corresponding photometric variability (over the same spectral band) and the projected rotation \citep{Aigrain2012,Vanderburg2016}. With photometric variability ranges from \citet{Garcia2014} and \citet{Santos2021} covering everything from a few hundred to a few $10^5$ ppm (and likely with a bias towards higher variability with increasing period) it is difficult to estimate what the expected intrinsic and projected variability should be for a star like \gamcepa.  
Our asteroseismic analysis did not allow us to place any strong constraints on the stellar inclination nor rotation \citep[as we did not see evidence of rotationally split oscillation modes;][]{Lund2014,Campante2016}, but is consistent with a low projected rotation.

The exact orientation of \gamcepa is very interesting in the context of the dynamic history of the system, where a pole-on configuration place some rather tight constraints on the tilt between the stellar spin axis of \gamcepa and the orbital plane of \planet, the so-called obliquity. With the constraint of the orbital inclination by \citet{Reffert2011} in the interval 3.7-15.5$^\circ$, \gamcepa and \planet would (if \gamcepa is seen pole-on) either be aligned or anti-aligned.

\subsection{Contemporaneous data}

The study we carried out here is in many ways similar to the one conducted by \citet{Arentoft2019} for the planet-hosting red giant $\epsilon$~Tauri, where data from the K2 mission \citep{Howell2014} were paired with SONG data. In \citet{Arentoft2019} they derived the amplitude difference of the oscillations between the space-based photometry and the ground-based RVs. However, as the K2 and SONG data were not collected simultaneously, the interpretation of the amplitude difference was slightly hampered. 

While we do not investigate the amplitude difference here, we do have simultaneous photometry and RVs from 2019. This will be used in a forthcoming paper (Lund et al., in prep.), which will also include SONG RVs obtained simultaneously with two additional \tess sectors, and in addition to amplitude differences, phase differences will also be investigated.

\section{Conclusions}\label{sec:conc}

Through long-term, high-cadence monitoring utilising both ground-based spectroscopic observations from the SONG network and space-borne photometry from \tess, we present an in-depth asteroseismic study of the planet-hosting, binary, RGB star \gamcepa. 

In our seismic analysis we obtained both the global seismic parameters, \numax and \dnu, and individual frequencies, which are tabulated in \tref{tab:freqs}. To provide additional constraints when modelling the frequencies, we performed a spectral analysis using data from the FIES spectrograph to obtain values for \teff, $\log g$, and $\rm [Fe/H]$, and we derived a luminosity using the distance and $G$-band magnitude from \gaia. We modelled the frequencies with the aforementioned constraints using BASTA. From BASTA we obtained a mass of \finalmass, a radius of \finalradius, and an age of \finalage.

We used our SONG RVs to fit the binary as well as the planetary orbit. Using literature values for the inclinations of the orbits and our derived orbital parameters with our mass for \gamcepa, we obtain masses of $M_{\rm B}=0.328^{+0.009}_{-0.012}$~M$_\odot$ and $M_{\rm Ab}=6.6^{+2.3}_{-2.8}$~M$_\oplus$ for \gamcepb and \planet, respectively.

Finally, amplitude and phase differences between SONG and \tess data will be analysed in a forthcoming paper (Lund et al., in prep.). This will be done using the simultaneous SONG and \tess data we have presented here as well as additional, simultaneous SONG and \tess data that span two \tess sectors. In this forthcoming analysis the improved frequency resolution will enable us to look further into the characteristics of individual modes, including line widths, amplitudes, and possible asymmetries.  

\begin{acknowledgements}
We thank the anonymous referee for comments and suggestions which improved the manuscript. 
All publicly available SONG data can be accessed through the SONG Data Archive (SODA; \url{https://soda.phys.au.dk/index.php}).
The \tess data are available through the Mikulski Archive for Space Telescopes (MAST; \url{https://archive.stsci.edu/missions-and-data/tess}).
The authors thank Earl P. Bellinger for help regarding modelling in the early stages of this work.
E.K. acknowledges the support from the Danish Council for Independent Research through a grant, No.2032-00230B.
Funding for the Stellar Astrophysics Centre is provided by The Danish National Research Foundation (Grant agreement no.: DNRF106).
%
This paper includes observations made with the SONG telescopes operated on the Spanish Observatorio del Teide (Tenerife) and at the Chinese Delingha Observatory (Qinghai) by Aarhus University, by the Instituto de Astrofísica de Canarias, and by the National Astronomical Observatories of China.
We acknowledge the use of public TESS data from pipelines at the TESS Science Office and at the TESS Science Processing Operations Center. Resources supporting this work were provided by the NASA High-End Computing (HEC) Program through the NASA Advanced Supercomputing (NAS) Division at Ames Research Center for the production of the SPOC data products.
M.V. acknowledge support from NASA grant 80NSSC18K1582.
This research made use of Astropy,\footnote{http://www.astropy.org} a community-developed core Python package for Astronomy \citep{astropy2013,astropy2018,astropy:2022}. 
This research made use of matplotlib \citep{matplotlib}.
The numerical results presented in this work were partly obtained at the Centre for Scientific Computing, Aarhus \url{https://phys.au.dk/forskning/faciliteter/cscaa/}.
This research has made use of the SIMBAD database,
operated at CDS, Strasbourg, France.
\end{acknowledgements}

%
%

\bibliographystyle{aa} 
\bibliography{bibliography} 

\appendix
\section{Additional figures}

\begin{figure*}
    \centering
        \includegraphics[width=\textwidth]{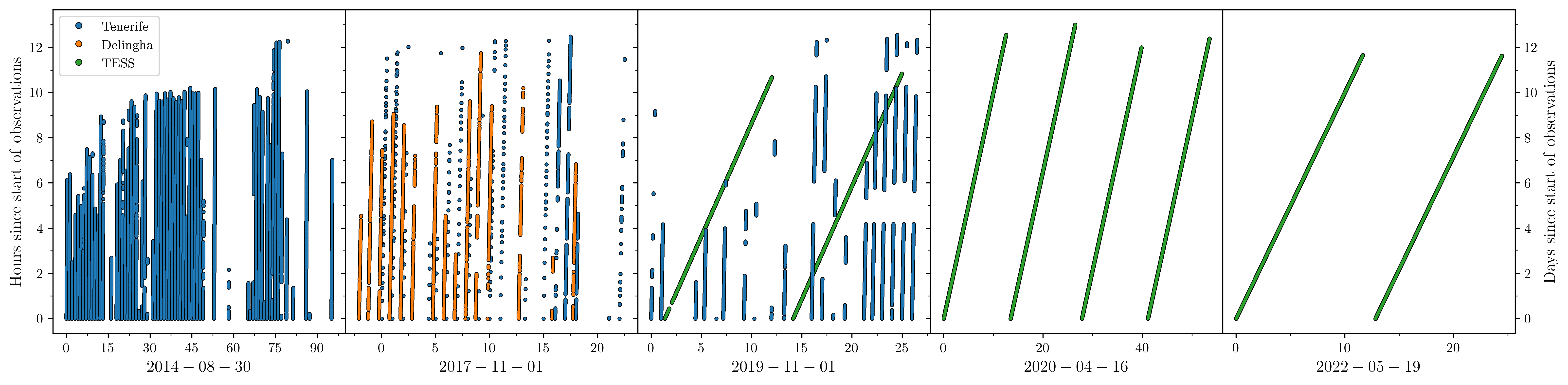}
    \caption{{\bf Timestamp-mod-timestamp plot.} In the first three panels we show the different SONG campaigns of \gamcep, where we have divided each dataset into 24~hr intervals and have plotted these intervals (in hours) against the time from the start of that campaign. The blue points are data from the Tenerife node, and the orange points are data from the Delingha node. In the two last panels we have plotted the TESS data as green points, where this time we have divided the data into intervals of 13.7~d (i.e. the orbital period of TESS), and again we have plotted these intervals (in days) against the start of a campaign. In panel three, the start of the campaign is taken to be the start of the SONG campaign, clearly showing the overlap between the 2019 SONG and TESS data, whereas the start of the campaign (covering two sectors) in the last panel is taken as the start of sector 24.}
    \label{fig:campaigns}
\end{figure*}

\begin{figure*}
    \centering
    \includegraphics[width=\textwidth]{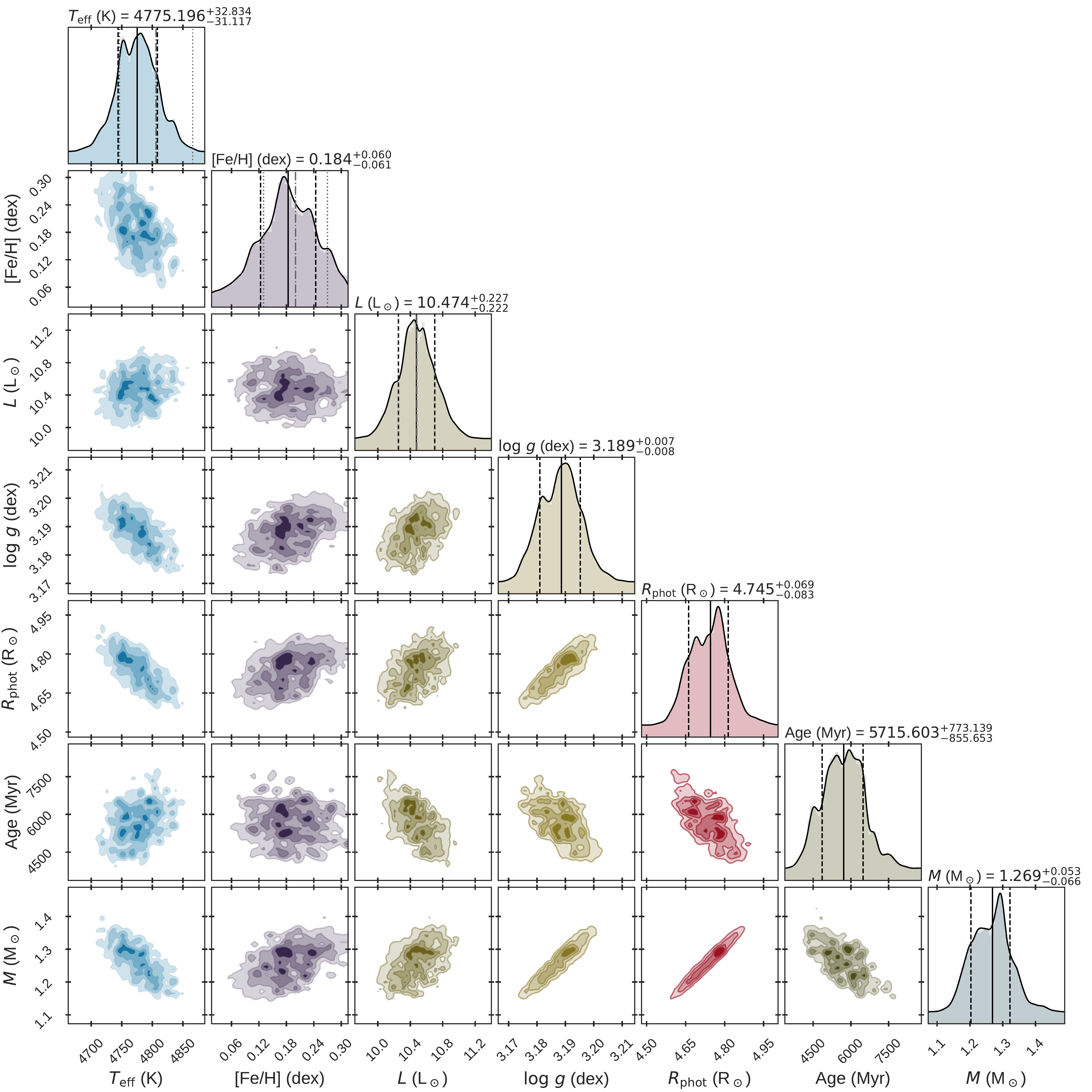}
    \caption{{\bf BASTA correlation plot} showing the resulting distributions from BASTA for the physical properties for \gamcepa.}
    \label{fig:corner}
\end{figure*}

\section{Additional table}

\begin{table}
    \centering
    \caption{{\bf Mode frequencies.}  
    }
    \begin{threeparttable}

    \begin{tabular}{c c c}
\toprule 
Order & Degree & Frequency \\  &  & ($\mu$Hz) \\ 
\midrule
8 & 2 & $132.5 \pm 0.4$ \\ 
9 & 0 & $134.3 \pm 0.3$ \\ 
9 & 2 & $146.0 \pm 0.4$ \\ 
10 & 0 & $148.1 \pm 0.2$ \\ 
10 & 2 & $160.29 \pm 0.13$ \\ 
11 & 0 & $162.23 \pm 0.18$ \\ 
11 & 1 & $169.4 \pm 0.2$ \\ 
11 & 2 & $174.69 \pm 0.13$ \\ 
12 & 0 & $176.53 \pm 0.18$ \\ 
12 & 1 & $183.1 \pm 0.2$ \\ 
13 & 1 & $184.33 \pm 0.13$ \\ 
12 & 2 & $188.90 \pm 0.11$ \\ 
13 & 0 & $190.69 \pm 0.13$ \\ 
14 & 1 & $197.7 \pm 0.3$ \\ 
15 & 1 & $198.89 \pm 0.14$ \\ 
13 & 2 & $203.37 \pm 0.15$ \\ 
14 & 0 & $205.16 \pm 0.19$ \\ 
16 & 1 & $211.72 \pm 0.20$ \\ 
17 & 1 & $213.07 \pm 0.20$ \\ 
14 & 2 & $217.6 \pm 0.4$ \\ 
15 & 0 & $219.77 \pm 0.17$ \\ 
18 & 1 & $226.9 \pm 0.2$ \\ 
15 & 2 & $232.8 \pm 0.2$ \\ 
16 & 0 & $234.4 \pm 0.2$ \\ 
19 & 1 & $240.9 \pm 0.3$ \\ 
16 & 2 & $247.3 \pm 0.3$ \\ 
17 \tnote{\textdagger} & 0 & $250.1 \pm 0.2$ \\ 
\bottomrule
    \end{tabular}
    \begin{tablenotes}
        \item The observed individual mode frequencies extracted from the product power spectrum.
        \item[\textdagger] Dubious mode (\sref{sec:freq_ex}).
    \end{tablenotes}
    \end{threeparttable}
    \label{tab:freqs}
\end{table}


\end{document}